\documentclass[]{aa}

\usepackage{txfonts}
\usepackage{float}
\usepackage{graphics}
\usepackage{subfigure}
\usepackage{longtable}
\usepackage{natbib}
\usepackage{epsfig}
\bibpunct{(}{)}{;}{a}{}{,} 

\newcommand{\rms}{r{.}m{.}s{.}~}

\newcommand{\eg}{e{.}g{.}~}

\newcommand{\fg}{Fig{.}~}
\newcommand{\fgs}{Figs{.}~}
\newcommand{\sct}{Sect{.}~}

\newcommand{\bt}{{\tt B/T}}
\newcommand{\arm}{{\tt Arm Strength}}
\newcommand{\curv}{{\tt Arm Curvature}}
\newcommand{\rot}{{\tt Arm Rotation}}
\newcommand{\bbar}{{\tt Bar Length}}
\newcommand{\iring}{{\tt Inner Ring}}
\newcommand{\oring}{{\tt Outer Ring}}
\newcommand{\pseu}{{\tt Pseudo-Ring}}
\newcommand{\pert}{{\tt Perturbation}}
\newcommand{\dust}{{\tt Visible Dust}}
\newcommand{\disp}{{\tt Dust Dispersion}}
\newcommand{\floc}{{\tt Flocculence}}
\newcommand{\spot}{{\tt Hot Spots}}
\newcommand{\incc}{{\tt Inclination-Elongation}}
\newcommand{\inc}{{\tt Incl-Elong}}
\newcommand{\cont}{{\tt Contamination}}
\newcommand{\mult}{{\tt Multiplicity}}

\begin{document}


  \title{The EFIGI catalogue of 4458 nearby galaxies with morphology \,\,\,\,\,
         II. Statistical properties along the Hubble sequence}
   \author{V. de Lapparent\inst{1,2}, A. Baillard\inst{1,2} \and E. Bertin\inst{1,2}}

  \institute{Universit\'e Pierre et Marie Curie-Paris 6, Institut d'Astrophysique de Paris, 98 bis Boulevard Arago, F-75014 Paris, France \and
    CNRS, UMR 7095, Institut d'Astrophysique de Paris, 98 bis Boulevard Arago, F-75014 Paris, France}

\titlerunning{Statistical properties along the Hubble sequence}
\authorrunning{de Lapparent, Bertin and Baillard}

   \date{Received December 31, 2010; accepted March 22, 2011}
  
\abstract
{}
{The EFIGI catalogue of 4458 galaxies extracted from the 
  PGC and SDSS DR4 was designed to provide a multiwavelength reference
  database of the morphological properties of nearby galaxies. The sample is
  limited in apparent diameter and densely samples all RC3 Hubble types.}
{We examine the statistics of the 16 EFIGI shape attributes, 
  describing the various dynamical components, the texture, and
  the contamination by the environment of each galaxy.  
  Using the redshifts from SDSS, HyperLeda, or NED
  for 99.53\% of EFIGI galaxies, we derive estimates of absolute major
  isophotal diameters and the corresponding mean
  surface brightness in the SDSS $g$-band.} 
{We study the variations of the EFIGI morphological attributes with  
  Hubble type and confirm that the visual Hubble sequence is a decreasing sequence of
  bulge-to-total ratio and an increasing sequence of disk contribution to the 
  total galaxy flux. There is, nevertheless, a total spread of approximately
  five types for a given bulge-to-total ratio, because the Hubble 
  sequence is primarily based on the strength and pitch angle of the spiral arms, 
  independently from the bulge-to-total ratio. A steep decrease in the
  presence of dust from Sb to Sbc-Sc types appears to produce the grand spiral design 
  of the Sc galaxies.
  In contrast, the scattered and giant HII 
  regions show different strength variation patterns, with peaks for types Scd and
  Sm; hence, they do not appear to directly participate in the 
  establishment of the visual Hubble sequence. The distortions from a symmetric profile
  also incidentally increase along the sequence. Bars and inner rings are frequent 
  and occur in 41\% and 25\% of the disk galaxies respectively. Outer rings are half as 
  frequent than inner rings, and outer pseudo-rings occur in 11\% of barred galaxies.  
  Finally, we find a smooth decrease in mean surface brightness and intrinsic size
  along the Hubble sequence. The largest galaxies are cD, ellipticals
  and Sab-Sbc intermediate spirals (20-50 kpc in $D_{25}$), whereas
  Sd and later spirals are nearly half as big. S0 are intermediate 
  in size (15-35 kpc in $D_{25}$), and irregulars, compact and dwarf
  ellipticals are confirmed as small objects (5-15 kpc in $D_{25}$).
  Dwarf spiral galaxies of type Sa to Scd are rare in the EFIGI catalogue,
  we only find two such objects.}
{The EFIGI sample provides us for the first time with a quantitative
  description of the visual Hubble sequence in terms of the specific
  morphological features of the various galaxy types. }


\maketitle
  

\section{Introduction                           \label{intro}}
  
The large variety of galaxy morphological types, their evolution with
redshift, their relationship with galaxy masses and star formation
rate, and their different spatial distribution with galaxy density
indicate that galaxy morphometry -- measuring the shape
parameters of galaxies -- is important for understanding galaxy
formation and evolution.

With the availability of deep and extensive digital surveys with
well-resolved galaxy images such as the Sloan Digital Sky Survey
(SDSS), the CFHTLS (Canada-France-Hawaii
Telescope Legacy Survey), and the coming LSST (Large Synoptic Survey
Telescope) and DES (Dark Energy Survey), as well as state-of-the-art automatic
software such as the last version of {\sc SExtractor} \citep{bertin96,bertin10},
astronomers are becoming able to perform automatic morphometry analyses 
for large statistical samples out to redshift $\sim1$.

Understanding evolution in galaxy morphology requires a detailed
knowledge of the present-day Universe. This so far relies on large
compilations from various galaxy catalogues made from photographic
plates, in which only the Hubble morphological type and some estimate
of size and isophote orientation are known for each object (see for
example the Principal Galaxy Catalogue, hereafter PGC, \citealt{pgc}).
Detailed studies of specific morphological features do exist, but they are
limited to some specific features within limited statistical samples or
limited regions of the sky \citep{kormendy79,buta95,naim97,buta06,buta07}.

\begin{table*}[t]
\caption{Definition of the EFIGI morphological attributes grouped by 
attribute type}
\label{attrib_def}
\begin{center}
\begin{tabular}{p{3.0cm}p{13cm}}
\hline  
\hline
Attribute type / name & Attribute definition \\

\hline
{\bf Bulge}         & \\
\quad {\tt B/T }      & Ratio of bulge luminosity over total galaxy luminosity \\

\hline
{\bf Spiral arms}   & \\
\quad {\tt Arm Strength}   & Strength of the spiral arms, in terms of flux fraction relative to the whole galaxy  \\
\quad {\tt Arm Curvature}  & Average curvature of the spiral arms \\
\quad {\tt Arm Rotation}   & Winding of the spiral pattern (0-1 for clockwise, 2 for no preferred direction, 3-4 for counterclockwise) \\

\hline
{\bf Dynamical features} & \\
\quad {\tt Bar Length}   & Length of central bar component  \\
\quad {\tt Inner Ring}  & Strength of inner ring, inner lens or inner pseudo-ring (located inside the disk
     and/or spiral arm pattern and near the end of the bar)  \\
\quad {\tt Outer Ring}  & Strength of outer ring (located outside the disk and/or spiral arm pattern)  \\
\quad {\tt Pseudo-Ring}    & Type and strength of outer pseudo-rings $R_1^\prime$, $R_2^\prime$ and $R_1R_2^\prime$ \\
\quad {\tt Perturbation}   & Deviation of the light distribution from a profile with rotational symmetry\\

\hline
\multicolumn{2}{l}{\bf Texture} \\
\quad {\tt Visible Dust}    & Strength of features tracing the presence of dust: obscuration and/or diffusion of star light by a dust lane or molecular clouds  \\
\quad {\tt Dust Dispersion} & Patchiness of the dust distribution (smooth and sharp lanes or strongly irregular patches) \\
\quad {\tt Flocculence}     & Flocculent aspect of the galaxy due to scattered HII regions  \\
\quad {\tt Hot Spots}       & Strength of regions with very high surface brightness, including giant regions of star formation, active nuclei, or stellar nuclei \\

\hline
{\bf Appearance}  & \\
\quad {\tt Inclination-} {\tt Elongation}  & For disk galaxies, inclination
     $f=1-\cos\theta$, where $\theta$ is the angle between the rotation axis of the disk and the light-of-sight of the observer, or equivalently between the galaxy disk
  and the plane of the sky (0$^\circ$ for face-on, 90$^\circ$ for edge-on)\\
 & For spheroidal galaxies, apparent elongation of the object $f = 1 - b/a$ (where $a$ and $b$ are the apparent major and minor axis lengths). \\
\quad {\tt Arm Rotation}   & see above, in ``Spiral arms'' attribute type\\               

\hline
{\bf Environment}   & \\
\quad {\tt Contamination}  & Severity of the contamination by bright stars, overlapping galaxies or image artifacts (diffraction spikes, star halos, satellite trails, electronic defects)  \\
\quad {\tt Multiplicity}   & Abundance of neighbouring galaxies differing by less than $\sim 5$ mag from the main galaxy and centred within 0.75 $D_{25}$ from its centre\\

\hline
\end{tabular}
\end{center}
\end{table*}
\begin{table*}
\caption{EFIGI morphological types}
\label{types}
\begin{center}
\begin{tabular}{lcccccccccccccccccc}
\hline  
\hline
Type & cE & E & cD & S0$^-$ & S0 & S0$^+$ & S0a & Sa & Sab & Sb & Sbc & Sc & Scd & Sd & Sdm & Sm & Im & dE \\
T & -6 & -5& -4 & -3     & -2 & -1     & 0   & 1  & 2   & 3  & 4   & 5  & 6   &  7 &   8 &  9 & 10 & 11 \\
\hline
\end{tabular}
\end{center}
\end{table*}

To obtain a reference sample for morphological studies, we
have designed the EFIGI (``Extraction de Formes 
Id\'ealis\'ees de Galaxies en Imagerie'') catalogue, a multiwavelength database
of 4458 galaxies with resolved digital images and detailed morphological
information. The sample is described in \citet{baillard08} and in
the companion article \citep[][ Paper I hereafter]{baillard11a}. 
The objects were selected from the PGC for their reliable RC3 morphological 
types \citep{rc3,rc3vizier} and for their inclusion
within the SDSS\footnote{http://www.sdss.org} DR4 photometric survey 
(photometric and spectroscopic data were  
obtained from the SDSS DR5 catalogue however). 
The resulting EFIGI database provides a large diversity of types and 
galaxy characteristics over the full 6670 deg$^2$ of the SDSS DR4. 
The EFIGI catalogue is publicly available and may be queried
from a web browser at {\tt http://www.efigi.org}, and from the 
``Centre de Donn\'ees Astronomiques de Strasbourg'' (CDS) using the
ViZiER Catalogue Service.

The EFIGI morphological description includes an updated RC3-based
Hubble type, and 16 shape attributes determined visually by 10
astronomers, which were subsequently homogenised. The attributes 
measure the significance of the different components of a galaxy
(bulge, spiral arms and other dynamical features, texture) 
as well as its appearance on the sky (inclination or elongation)
and its environment (contamination and multiplicity).
Paper I describes the various statistical tests that demonstrate the
reliability of this morphological description, and reports on
the properties of the catalogue in terms of sky coverage, clustering, 
galaxy counts, magnitude completeness, and morphological fractions.
The variety of the EFIGI database provides a unique description of
the local Universe and allows one to perform quantitative
statistical analyses over a large variety of morphological properties
at $z\la0.05$. 

In \sct\ref{cfigi} we briefly describe the EFIGI attributes, the
redshifts and magnitudes used in the present analysis, and the catalogue
diameter and magnitude limits. We then examine
in \sct\ref{morph} the statistics of the various EFIGI attributes as a
function of morphological type. We show that they provide a detailed
and quantitative description of the Hubble sequence. Finally, we
examine in \sct\ref{diam} the variations in apparent diameter and mean
surface brightness of the EFIGI galaxies as a function of morphological
type.


\section{The EFIGI catalogue                  \label{cfigi}}


\subsection{Morphological attributes and types        \label{attrib}}

The EFIGI Hubble type and 16 morphological attributes were measured
by visual examination of the composite ``$gri$'' colour image of
each galaxy, derived from the SDSS FITS images using the
{\it AstrOmatic} software.\footnote{\tt http://www.astromatic.net}
Table \ref{attrib_def} lists the definition of the 16 attributes.
Each attribute can take five possible values and has a lower and upper
confidence limit. Note that we here convert the original
decimal values ranging from 0 to 1 in Paper I
into integer values ranging from 0 to 4. We also use 
the inverse square of the confidence interval to weight
the attributes. This interval is defined
as the difference between the upper and lower limit plus 1, and
thus takes values between 1 and 5. Throughout 
the article, we refer to attribute values 1, 2, 3, and 4 by ``weak'', 
``moderate'', ``strong'', and ``very strong'' respectively, except for bars,
which we describe as ``short'', ``intermediate'', ``long'', and `` very long''. 

The EFIGI morphological sequence is based on the RC3 ``Revised Hubble
Sequence'' (RHS hereafter), and we call it the ``EFIGI Morphological
Sequence'' (EMS hereafter). The major difference between the two
sequences is that in the RHS non-magellanic irregulars (I0 type) are
not considered as a separate type in the EMS, but as galaxies of some
type of the EMS which undergo distortions in their profile measured
by the \pert\ attribute. Moreover, the EMS contains one additional
type, gathering the dwarf elliptical, dwarf lenticular, and dwarf
spheroidal galaxies, whereas they are classified as ellipticals in the
RHS.  Finally, the different ellongation stages of elliptical galaxies
in the RHS are not distinguished in the EMS, because this is measured
by the EFIGI \incc\ attribute. The various EFIGI morphological types
(hereafter ``EM-types'') of the EMS are listed in Table \ref{types}.

\begin{figure}
\resizebox{\hsize}{!}{\includegraphics{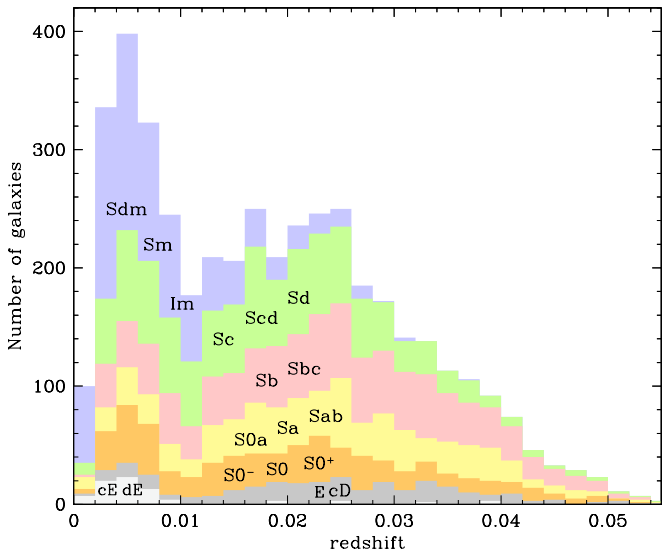}}
\caption{Redshift distribution of all EFIGI galaxies separated by
EFIGI morphological type, grouped as follows: cE-dE (light grey) ; E-cD
(grey); S0$^-$-S0-S0$^+$ (orange); S0a-Sa-Sab (yellow); Sb-Sbc (pink);
Sc-Scd-Sd (green); Sdm-Sm-Im (blue).}  
\label{z_hist}
\end{figure}


\subsection{Redshifts                   \label{z}}

We completed the EFIGI catalogue with the different measures 
of redshifts
extracted from the PGC \citep{pgc}, HyperLeda \citep{paturel2}, NED, and SDSS. 
The vast majority of EFIGI galaxies (4415) have a
HyperLeda redshift measurement. Many also have a NED redshift (4404). 
The advantage of the 
HyperLeda database is that it provides redshifts corrected for 
Virgocentric infall, which is particularly important for nearby galaxies
such as in the EFIGI catalogue. When the HyperLeda and NED heliocentric
redshifts differ by more than 0.0001 (which is about a 3-sigma difference
in both types of redshift, as well as the typical uncertainty of
SDSS redshifts), we adopted the NED heliocentric redshift, which shows 
the individual redshift values and allows one to trace the sources of 
discrepancies. 

As a result, we use the HyperLeda heliocentric redshift for 4079
of EFIGI galaxies, the NED redshift for 349 galaxies, and the SDSS redshift
for 9 galaxies (there is no need to use the PGC redshifts).
The SDSS and PGC redshifts are very useful however to check
the consistency with the HyperLeda and NED redshifts, and to distinguish among
both values when they differ; this also allowed us to discard some rare
erroneous HyperLeda and NED redshifts. 

In total, 4437 EFIGI galaxies among 4458 have an heliocentric
redshift, corresponding to a redshift completeness of 99.53\%.
When a catalogue
other than HyperLeda is used to assign an heliocentric redshift to an
object, we adopted the HyperLeda correction for virgocentric
infall if the adopted redshift differs from the HyperLeda value by less
than 0.0005. This yields 4411 galaxies with a redshift corrected for 
Virgocentric infall.

Twelve galaxies have a
negative HyperLeda redshift, even after correction for virgocentric
infall. Because these galaxies are very nearby (velocities between -8 km/s
and -350 km/s), their redshifts cannot be used as an estimate
of distance, and we discarded them.

\fg\ref{z_hist} shows
the resulting redshift distribution after the separation of EFIGI galaxies
into groups of two or three morphological types. The strong excess at
$z\sim0.005$ is caused by the Virgo cluster. There are only 72 EFIGI
galaxies beyond $z=0.05$, because the RC3 catalogue mainly contains 
galaxies with recession velocities lower than the corresponding value
of 15 000 km/s, that is, below a luminosity distance of $\sim 220$ Mpc.
Throughout the article, we convert redshifts into distances
using a Hubble constant $H_0=70$ km/s/Mpc \citep{freedman01}, and the 
currently standard cosmological parameters 
$\Omega_\mathrm{m}=0.3$ and $\Omega_\Lambda=0.7$ \citep{wmap5}. 

\fg\ref{z_hist} shows that the various morphological types of giant
galaxies - cD, E, S0 and spirals earlier than Sd (see \sct\ref{d25}) - are detected at all
redshifts in the EFIGI catalogue (although cD galaxies are too few to 
be sampled in the whole redshift range). 
The intrinsically smaller and fainter objects - Sdm, Sm,
Im, cE, dE - are detected preferentially at lower redshifts, owing to the
apparent diameter limit of the RC3. This is discussed in more detail in 
\sct\ref{lim} below.

\begin{figure}
\resizebox{\hsize}{!}{\includegraphics{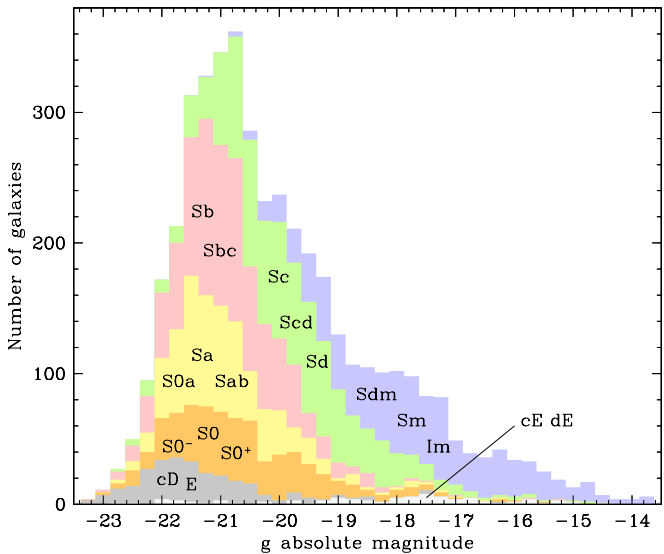}}
\caption{Absolute magnitude distribution in the $g$-band for all
EFIGI galaxies separated by EFIGI morphological type, grouped as
follows: cE-dE (light grey) ; E-cD (grey); S0$^-$-S0-S0$^+$ (orange);
S0a-Sa-Sab (yellow); Sb-Sbc (pink); Sc-Scd-Sd (green); Sdm-Sm-Im
(blue).}  
\label{mag_hist}
\end{figure}


\subsection{Photometry                   \label{phot}}

Because the SDSS photometric pipeline was designed for galaxies smaller
than 2 arcmin in diameter, whereas the SDSS contains many larger galaxies
(up to several degrees in isophotal diameter; see \fg\ref{d25_g}),
the pipeline fails for large galaxies, which are truncated
into several units. We showed in \fg 24 of Paper I that there 
can be as much as an 8-magnitude difference for some objects, mostly late-type 
galaxies. 

de Lapparent \& Bertin (2011a, in prep{.]) have thus recalculated the magnitudes of all 
EFIGI galaxies from the SDSS $ugriz$ images using the new version of 
{\sc SExtractor} \citep{bertin96,bertin10}, with the option to fit a bulge+disk 
PSF-convolved profile. The resulting apparent magnitudes in the SDSS 
\textit{ugriz} bands (see \fg 1 of de Lapparent \& Bertin 2011a, in prep{.}) confirm the 
disappearance of the spurious tail at faint apparent magnitude,
which is seen when using the SDSS photometry (in \fg 23 of Paper I).
The analysis of de Lapparent \& Bertin (2011a, in prep{.]) also shows that the SDSS DR5 fluxes
undergo losses of 0.5 to 2 magnitudes for bright galaxies ($9\le g\le16$),
owing to an overestimate of the sky background around large objects;
an improved sky background procedure has been implemented in the SDSS DR8,
\citealt{blanton11}.

The apparent magnitudes and luminosity distances are used to derive
absolute magnitudes for all EFIGI galaxies.  In this calculation, we use 
the k-corrections provided by the VAGC DR4 ``low-z'' catalogue \citep{vagc,blanton07}
for 1725 EFIGI galaxies to derive linear estimates of the k-corrections as a 
function of redshift and in each filter for the remaining objects; 
the whole process is described in detail in de Lapparent \& Bertin (2011a, in prep{.]).

The resulting distributions of absolute magnitudes in the SDSS
$g$-band (the most sensitive SDSS band) are shown in
\fg\ref{mag_hist} after separating EFIGI galaxies into groups of two or three
morphological types.  These curves result from
the combination of the apparent magnitude distributions, the
luminosity functions of the different galaxy types
\citep{lapparent03}, and the selection function of the EFIGI
catalogue (see \sct\ref{lim}). The graph
shows that the EFIGI catalogue densely samples the various types over a wide
absolute magnitude interval, including both the bright ends of the
lenticular and spiral galaxy distributions, and the faint ends of the
late spirals and irregulars.


\subsection{A diameter-limited catalogue                \label{lim}}
  
\begin{figure}
\resizebox{\hsize}{!}{\includegraphics{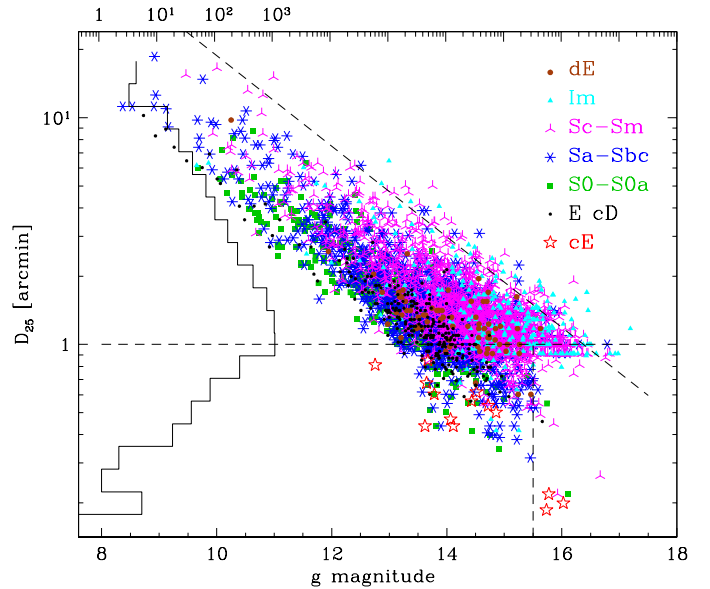}}
\caption{Distribution of apparent major isophotal diameter
  $D_{25}$ converted into arcminutes as a function of $g$-magnitudes
  measured by de Lapparent \& Bertin (2011a, in prep{.}) for the 4204 EFIGI galaxies for which
  these parameters are defined. Different symbols and colours are used for
  different EM-types. The histogram of $D_{25}$ in intervals of 0.1 arcminutes,
  overplotted vertically on a $\log$ scale and with corresponding labels 
  on the top axis of the graph, demonstrates the strong incompleteness 
  in the catalogue for $D_{25}<1'$ and $B_T>15.5$, materialised by the horizontal 
  and vertical dashed lines. The diagonal dashed line corresponds to a 
  surface brightness limit of 25 mag/arcsec$^2$. The EFIGI catalogue is 
  most complete between these diameter and surface brightness limits,
  and the crossing of these two lines 
  causes the disappearance of data points fainter than $g\sim16.5$.}
\label{d25_g}
\end{figure}
\begin{figure}
\resizebox{\hsize}{!}{\includegraphics{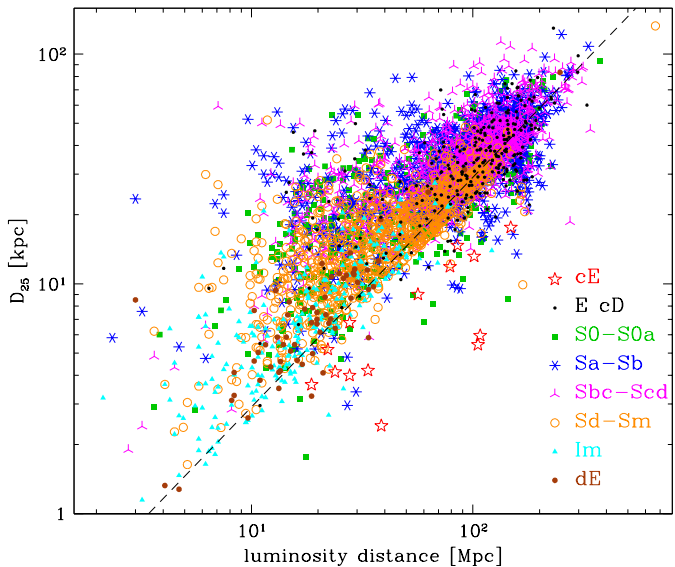}}
\caption{Distribution of apparent major isophotal diameter
  $D_{25}$ converted into kpc as a function of $g$ magnitudes, 
  for the 4104 EFIGI galaxies for which these
  parameters are defined. Different symbols and colours are used for different
  EM-types (beware of the different spiral grouping from that in \fg\ref{d25_g}).
  The $D_{25}<1'$ limit is plotted as a dashed line. Galaxies below this 
  diameter limit extend the minimum detectable diameter for 
  a given distance to a lower value.}
\label{d25_dl}
\end{figure}

Because the EFIGI sample was restricted to PGC galaxies having a
reliable RC3 measurement of Hubble type, the sample undergoes 
the selection effects of the RC3 \citep{rc3}.
This latter catalogue was aimed at being reasonably complete for galaxies
having an apparent diameter $D_{25}$\footnote{defined in the RC2 
system \citep{rc2} as the apparent major isophotal diameter measured 
at or reduced to a surface brightness level of 25.0
B/mag$^2$, in units of 0.1 arcminute.} larger than 1 arcminute,
a total $B_T$ magnitude brighter than about 15.5, and a
recession velocity not in excess of 15 000km/s. Half of the
RC3 objects however are galaxies that satisfy only the diameter
or magnitude condition, and in addition may have velocities higher
than 15 000 km/s.  

The resulting selection effects are shown in \fg\ref{d25_g}, where we plot 
the distribution of $D_{25}$ apparent diameters converted into arcminutes,
as a function of the $g$-band magnitudes measured by de Lapparent \& Bertin (2011a, in prep{.]).
The overplotted vertical histogram of $D_{25}$ shows strong incompleteness for 
$D_{25}<1$ arcmin, below the horizontal dashed line, where there are 1204
galaxies, which is more than one quarter of the full EFIGI catalogue. 
 This allows the inclusion in the sample of galaxies with smaller 
diameter than the rest of the objects, like cE galaxies, some of which reach
$D_{25}\simeq0.2$ arcmin (and at $g>15.5$). 
Following the RC3 selection, almost all of the galaxies with $D_{25}<1$ arcmin
are brighter than $g\sim15.5$, as indicated by the vertical dashed line.
Moreover, we find that for each Hubble type (including 
cD and dE), the galaxies with $D_{25}<1$ arcmin (hence $g<15.5$) represent a 
comparable fraction of 1/3 to 1/4 of the objects.
 
Finally, a diagonal dashed line shows the 25 mag/arcsec$^2$ limit in 
surface brightness, below which lie most EFIGI galaxies, except a few Sd-Sm 
and Im types (this is best seen in \fg\ref{d25_sb25}, \sct\ref{sb}).
This surface brightness limit corresponds to the level below which it is 
difficult to visually detect isophotes or entire objects from a paper copy
of the Palomar Observatory Sky Survey plates \citep{poss59}.
We therefore expect a high completeness rate for the EFIGI sample above
this surface brightness limit, that is, for $D_{25}$ below the diagonal line
in \fg\ref{d25_g}.

The diameter limit of the RC3 results in a distance-dependent limit in
the physical size of EFIGI galaxies. We estimated the
intrinsic diameter of galaxies by multiplying the apparent diameter 
$D_{25}$ by the angular-diameter distance. This is applicable to a subsample of
4152 EFIGI galaxies: among the total number of 4458 EFIGI galaxies, 257 have no
measure of $D_{25}$ in the RC3, and 49 additional galaxies have 
no measure of redshift. We show in \fg\ref{d25_dl} the resulting variations
in the intrinsic absolute major diameter as a function of luminosity distance
for the EFIGI catalogue. The incompleteness at $D_{25}<1$ arcmin in
\fg\ref{d25_g} results in a minimal detectable diameter for each distance,
materialised as a diagonal line in \fg\ref{d25_dl}. The 1204 EFIGI galaxies 
located below this diameter limit make up a sample
of intrinsically smaller galaxies of all Hubble types: 
for example cE galaxies, which can be detected up to large distances in the EFIGI 
catalogue ($\sim200$ Mpc). 
We notice that the various types are located at different positions
along the $D_{25}=1$ arcmin line, according to their range of absolute
magnitude.  

Another selection effect that adds on the RC3 apparent diameter 
limit is the following: over the area of sky covered by
the SDSS photometric DR4 survey, the EFIGI sample contains all PGC
galaxies having an RC3 morphological classification based on several
measurements to guarantee a reliable RC3 type (see Paper I). 
This effect implies that fewer high inclination spiral galaxies are selected, 
because of the poorer visibility of their morphological features
(see \sct\ref{env-app}). This selection effect can only be evaluated by 
comparison to a complete census of galaxies brighter than 
$g\sim17$ and with reliable photometry; this is not provided by the SDSS
because the various releases contain a large number of spurious sources and 
the photometry pipeline fails for large objects (Paper I).
As a result, the galaxy groups, clusters and large-scale structures contained
in the EFIGI catalogue might not be sampled in proportion to their true spatial 
density. The more than $\approx 80$\% completeness of the EFIGI catalogue for 
galaxies with $10<g<14$ (Paper I) limits the 
amplitude of this ``density bias'' at bright magnitudes, however.

Interestingly, because of the various selection effects at play in the EFIGI 
sample, the number of EFIGI galaxies per morphological type undergoes a limited 
spread after excluding the intrinsically rare types (cE, cD, dE). The median 
number of galaxies per type is $\sim 257$, corresponding to Sa galaxies,
the lowest value is $152$ for S0$^+$, and the four highest values are 
$518$ for Sb, $472$ for Sbc, $445$ for Scd, and $355$ for Sdm; 
all remaining types have numbers of galaxies that deviate from the
median by less than 50\% of its value.  
This constitutes a major difference between the EFIGI catalogue and the other
recent morphological follow-up studies  \citep{nair10,lintott08}, 
which are based on apparent magnitude-limited subsamples of the SDSS.
As shown in Paper I, the EFIGI catalogue largely oversamples late 
spirals (Sd-Sdm-Sm) and irregulars compared to magnitude-limited surveys.
Part of these excess late-type galaxies in the EFIGI catalogue are located 
in the small diameter extension of \fg\ref{d25_g}, at $D_{25}<1$ arcmin and 
$g\le15.5$.


\section{Morphological characteristics along the Hubble sequence \label{morph}}

\subsection{Environment and appearance                      \label{env-app}}

Before examining the EFIGI morphological attributes {\it per se}, that is,
those which contribute to the description of the morphological type, we
examine the ``environment'' and ``appearance'' EFIGI attributes, which
are independent of the internal galaxy properties for \incc\ and
\cont, and only indirectly related for \mult\ (via the morphology-density 
relation, \eg\ \citealt{blanton09}). We also examine here the \rot\
attribute, which can also be considered an ``appearance'' attribute,
because the direction of winding of the spiral arms depends on the 
orientation of the galaxy with respect to the line-of-sight.

Table \ref{attrib_stat_env} shows the fraction of galaxies with
no, weak, moderate, strong, or very strong \cont\ and \mult; attribute 
values with less than $\sim10$\% galaxies have been grouped together 
(1 and 2 for \mult, and 3 and 4 for \cont\ and \mult). The listed 
fractions show no dependence of \cont\ on EM-type for types S0$^-$ and later.
Compared to the other types, there is a $\sim 10$\% deficiency in cD and cE-E 
galaxies with \cont\ = 0, and a corresponding excess of objects with
\cont\ = 2. This is probably because of the large envelopes
of E, and even more so cD galaxies, which increases the probability of
contamination by any type of real objects (stars or galaxies), or
by artefacts. 

In all tables of attribute statistics listed in the article, the
quoted error bars are the Poisson uncertainties in the considered
fractions of objects.  For the large part of lower and upper
confidence limits equal to the attribute value plus and minus one
level, that is a confidence interval of 3, these uncertainty
estimates are adequate.  For the attribute values that are equal to
one or both of their lower or upper confidence limit, that is, a 
confidence interval of 1 or 2, we choose to be
conservative by ignoring the reduction in uncertainty from the Poisson
estimates.  Listed fractions and uncertainties are rounded to the
closest percent, and uncertainties in the interval 0.1 to 0.5\% are
rounded to 1\%.

\begin{table}
\caption{Statistics of the EFIGI ``Environment'' attributes given as a percentage of galaxies with given attribute values}
\label{attrib_stat_env}
\begin{center}
\begin{tabular}{p{1.5cm}p{0.5cm}p{0.6cm}p{0.6cm}p{0.5cm}p{0.6cm}p{0.6cm}p{0.8cm}}  
\hline  
\hline
EFIGI Type & \multicolumn{4}{c}{\tt Contamination} & \multicolumn{3}{c}{\tt Multiplicity}\\
                 &   0  &  1  & 2 &  3-4 &   0  &  1-2 & 3-4 \\
\hline
              cD & 17$\pm$7 & 43$\pm$12 &  35$\pm$10 & 4$\pm$3 & 46$\pm$12 & 39$\pm$11 & 15$\pm$6\\  
            cE E & 16$\pm$3 & 52$\pm$6 &  24$\pm$4 & 8$\pm$2 & 74$\pm$7  & 26$\pm$4 &  $\quad$- \\  
S0$^-$ S0 S0$^+$ & 26$\pm$3 & 52$\pm$4 &  17$\pm$2 & 5$\pm$1 & 82$\pm$5  & 17$\pm$2 &  1$\pm$ 1 \\ 
          S0a Sa & 28$\pm$3 & 53$\pm$4 &  15$\pm$2 & 4$\pm$1 & 86$\pm$6 & 14$\pm$2 &  0$\pm$ 1 \\  
          Sab Sb & 34$\pm$3 & 49$\pm$3 &  13$\pm$1 & 4$\pm$1 & 86$\pm$5 & 14$\pm$2 &  0$\pm$ 1 \\  
          Sbc Sc & 35$\pm$3 & 50$\pm$3 &  11$\pm$1 & 4$\pm$1 & 91$\pm$5 & $\;$9$\pm$1 &  0$\pm$ 1 \\  
          Scd Sd & 37$\pm$3 & 50$\pm$3 &  10$\pm$1 & 3$\pm$1 & 93$\pm$5 & $\;$7$\pm$1 &  0$\pm$ 1 \\  
          Sdm Sm & 34$\pm$3 & 52$\pm$4 &  11$\pm$1 & 4$\pm$1 & 93$\pm$5 & $\;$7$\pm$1 &  0$\pm$ 1 \\ 
           Im dE & 27$\pm$3 & 53$\pm$5 &  13$\pm$2 & 6$\pm$2 & 88$\pm$7 & 12$\pm$2 &  0$\pm$ 1 \\ 
\hline
All types        & 31$\pm$1 & 51$\pm$1 &  13$\pm$1 & 4$\pm$1 & 88$\pm$2 & 12$\pm$1 &  1$\pm$ 1 \\
\hline
\end{tabular}
\end{center}
{\it Notes:}
Null fractions are replaced by ``-'' for clarity.
Fractions are grouped by types having similar values of the considered attribute.
\end{table}
\begin{table}
\caption{Statistics of the EFIGI ``Appearance'' attributes given as a percentage of galaxies with given attribute values}
\label{attrib_stat_pres}
\begin{center}
\begin{tabular}{lrrrrr}
\hline  
\hline
EFIGI Type & \multicolumn{5}{c}{\tt Inclination/Elongation} \\
           &   0   &  1    &  2    &   3   &  4   \\  
\hline
  cE E &          30$\pm$ 4 & 51$\pm$ 6 & 19$\pm$ 3 &        -  &        - \\ 
   cD  &          11$\pm$ 5 & 48$\pm$12 & 39$\pm$11 & 2$\pm$ 2 &        - \\ 
S0$^-$S0 S0$^+$ & 14$\pm$ 2 & 27$\pm$ 3 & 30$\pm$ 3 & 14$\pm$ 2 & 15$\pm$ 2 \\ 
 S0a Sa  &        16$\pm$ 2 & 25$\pm$ 3 & 30$\pm$ 3 & 21$\pm$ 2 &  9$\pm$ 2 \\ 
 Sab Sb  &        16$\pm$ 2 & 33$\pm$ 2 & 26$\pm$ 2 & 16$\pm$ 2 & 10$\pm$ 1 \\ 
 Sbc Sc  &        16$\pm$ 2 & 24$\pm$ 2 & 27$\pm$ 2 & 17$\pm$ 2 & 16$\pm$ 2 \\ 
 Scd Sd  &        18$\pm$ 2 & 21$\pm$ 2 & 14$\pm$ 2 & 17$\pm$ 2 & 29$\pm$ 2 \\ 
 Sdm Sm  &        21$\pm$ 2 & 27$\pm$ 2 & 22$\pm$ 2 & 17$\pm$ 2 & 15$\pm$ 2 \\ 
 Im dE   &         5$\pm$ 1 & 48$\pm$ 5 & 22$\pm$ 3 & 15$\pm$ 2 & 10$\pm$ 2 \\ 
\hline
   All disks &    17$\pm$ 1 & 26$\pm$ 1 & 24$\pm$ 1 & 17$\pm$ 1 & 16$\pm$ 1 \\ 
\hline
                  & \multicolumn{3}{c}{\tt Arm rotation}\\
                  &  0-1 &  2   &  3-4 \\  
\hline
   S0$^-$ S0      & 1$\pm$1     &  98$\pm$ 7 &  1$\pm$ 1 \\ 
  S0$^+$ S0a      &  7$\pm$ 2    &  87$\pm$ 7 &  6$\pm$ 1 \\ 
      Sa          & 20$\pm$ 3    &  58$\pm$ 6 & 23$\pm$ 3 \\ 
 Sab Sb Sbc       & 39$\pm$ 2    &  28$\pm$ 2 & 34$\pm$ 2 \\ 
  Sc Scd Sd       & 31$\pm$ 2    &  41$\pm$ 2 & 28$\pm$ 2 \\ 
      Sdm Sm      & 12$\pm$ 2    &  74$\pm$ 5 & 13$\pm$ 2 \\ 
      Im          & 1$\pm$ 1     &  98$\pm$ 9 &  1$\pm$ 1 \\ 
      dE          & 1$\pm$ 1     & 97$\pm$17 &  1$\pm$ 1 \\ 
\hline
   All disks      & 24$\pm$ 1    &  53$\pm$ 2 & 22$\pm$ 1 \\ 
\hline
\end{tabular}
\end{center}
{\it Notes:}
Null fractions are replaced by ``-'' for clarity.
Fractions are grouped by types having similar values of {\tt Inclination/Elongation} and \rot\
  attributes.
``All disks'' corresponds to types from S0$^-$ to Sm.
For cE, E and cD galaxies, the \rot\ attribute is undefined because none 
of these galaxies have spiral arms. And very few of the S0$^-$, S0, Im and dE have
detectable spiral arms.
\end{table}

The statistics for the \mult\ attribute also show that spirals, 
Im and dE galaxies have
similar \mult\ distributions values (82-93\% with attribute
value 0, and 7-14\% with values 1-2), whereas E, cE and S0 galaxies
undergo a higher rate of weak \mult\ (74-82\% with attribute
value 0, and 26-17\% with values 1-2), owing to their large envelopes.
The combination of the richer environment of clusters of
galaxies and their higher density in cD, E, S0, Im and dE galaxies
also increases the probability of overlapping by another galaxy, hence
the higher percentages of objects with \mult\ = 1-2 for these
types. The very extended envelopes of cD galaxies explain why
there is an even more drastic increase in their fractions
of objects with weak and strong \mult\ ($15\pm6$\% for attribute values 3-4,
$39\pm11$\% for 1-2 attribute values), and nearly a factor of 2 decrease in
the fraction of galaxies with null attribute value: cD galaxies are 
located in the potential wells of rich galaxy clusters, where the galaxy 
density is the highest, hence the presence of neighbouring galaxies is high.

Table \ref{attrib_stat_pres} shows the distribution of the \incc\ 
attribute (\inc\ hereafter) grouped by types with similar values of
the attribute.  
As expected, none of the pure-bulge galaxies (cE, E, and cD) have an
\inc\ attribute value of 4, because this high value can only occur for
highly inclined disks; and only a few cD galaxies have an attribute
value of 3. Note however the apparent systematically higher elongation of
cD galaxies compared to cE and E in the EFIGI catalogue.  This may be
a true physical effect, caused by the dissipationless collapse of
aspherical mass concentrations and its impact on first-ranked
galaxies in clusters \citep{rhee90}.  

The other statistics in Table 
\ref{attrib_stat_pres} show the wide variety of disk inclinations for
the selected EFIGI lenticular and spiral galaxies, whose mean statistics are
indicated in the last line labelled ``All disks''. We emphasize that
the fewer disk galaxies selected at high \inc\ values (3-4)
in no way reflect any preferential orientation 
of galaxies on the sky, but are probably due to the less reliable RC3 
types for these objects, hence their under-representation in the EFIGI 
sample. There is also a diversity of elongations for Im and dE galaxies,
with a noticeable deficiency in round objects compared to spirals 
(for \inc\ = 0).

Finally, Table \ref{attrib_stat_pres} shows the distribution of the
\rot\ attribute, measuring the winding direction of the spiral
arms: an approximately symmetric distribution about the no-rotation
value of 2 is measured for all types, within the dispersion caused by
the number of galaxies in each EM class. As expected, only very few
(and weak) spiral arms are visible in S0$^-$, S0, Im and dE galaxies,
hence the small fractions with attribute values different from 2.
When averaging over all disk galaxies, the fractions of galaxies 
with the arm winding clockwise (\rot\ = 0 or 1) and counterclockwise 
(\rot = 3 or 4) agree to within $\sim1.5\sigma$, showing no
systematic effect in the EFIGI sample.

Below, we examine in detail the remaining 12 EFIGI
attributes, which each characterise a specific morphological feature
of galaxies, and hence contribute to the determination of the Hubble
sequence.


\subsection{Bulge-to-total ratio                       \label{bt}}
  
\begin{figure}
\resizebox{\hsize}{!}{\includegraphics{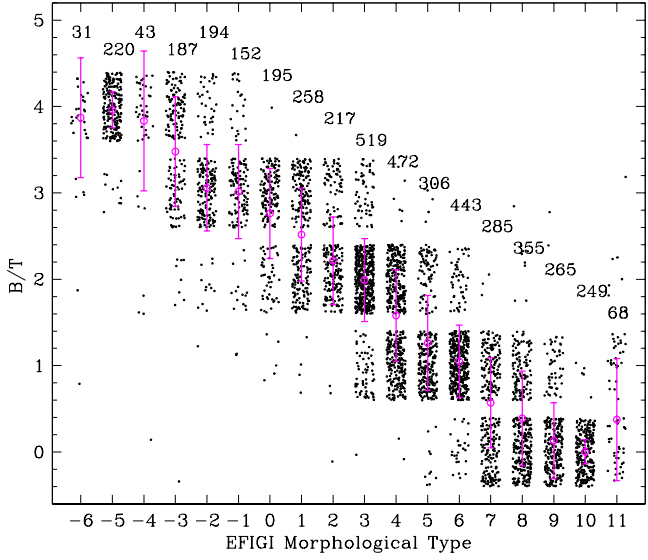}}
\caption{Distribution of \bt\ attribute versus the EFIGI morphological type for the
4458 galaxies in the EFIGI catalogue (see Table \ref{types} for the 
correspondence with types). In order to see the relative density of data points, 
they are spread in the horizontal and vertical direction, hence the apparent
rectangles. The weighted mean and weighted \rms\ dispersion in \bt\ are also
plotted for each type along with the corresponding number of galaxies. This
graph shows the strong correlation of morphological type with \bt.}
\label{emt_bt}
\end{figure}
\begin{figure}
\resizebox{\hsize}{!}{\includegraphics{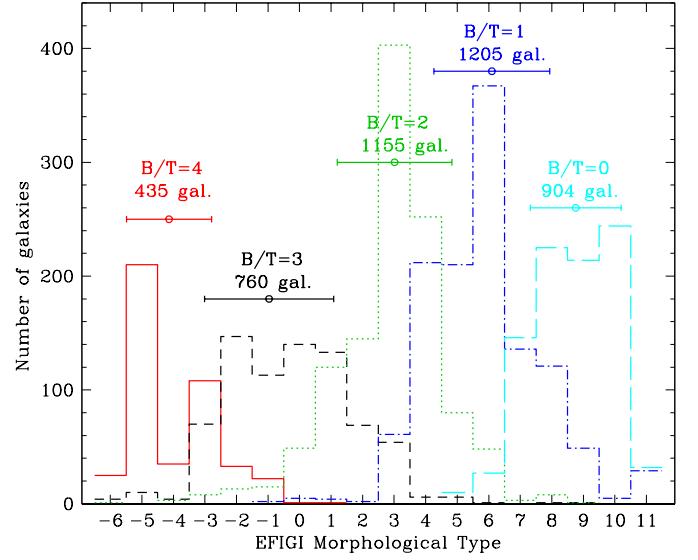}}
\caption{Histograms of EFIGI morphological type for the five values of \bt\
(decreasing from left to right). The weighted mean and \rms dispersion in
the morphological type are plotted above each histogram along with
the corresponding number of galaxies.  This graph shows the large
dispersion in morphological type for a given \bt.}  
\label{emt_bt_hist}
\end{figure}

The attribute measuring the ratio of light from the bulge over the total light of a galaxy,
denoted \bt, is tightly correlated with the morphological type, as shown
in \fg\ref{emt_bt} (the correspondence between the EM-types and their
ID number is listed in Table \ref{types}). There is a large
dispersion of EM-type for each value of \bt, however.  This is best seen in
\fg\ref{emt_bt_hist}, where we plot the histograms of EM-type for each
of the five values of \bt: the open circles and horizontal error bars show
the weighted mean and $\pm1\sigma$ dispersion within each \bt\ class, with
$\sigma$= 1.4, 1.8, 2.0, 1.9, and 1.3 for \bt\ = 0, 1, 2, 3, and
4 respectively. These dispersion values cannot be caused only by the EFIGI uncertainties
in the morphological type, because one third of EFIGI galaxies have a half 
confidence interval of 0.5, another third of 1, and 80\% of the last third of 
1.5.

In \fg\ref{emt_bt} we overplot the weighted mean value and $\pm1\sigma$
dispersion in the \bt\ attribute for each of the 18 EM-types. The dispersion
has the lowest values for types E (0.41) and Im (0.28), and the
highest value for types S0$^-$ (1.26) and dE (1.41). For the latter,
the large dispersion is because nearly half of the dE galaxies are 
nucleated \citep{binggeli91}, hence an attribute value of 1 or 2 has been
given to these objects (the \bt\ attribute value of 3 corresponds to PGC0009283, which
has a central hot spot with an attribute value of 2). For the S0, S0$^+$ and
spiral galaxies, the $\pm1\sigma$ dispersion in the \bt\ attrbute lies in the tight
interval between 0.84 for Scd and 1.10 for Sc, with a median value of
1.04.  These values are nearly unchanged when using only the 3106 galaxies 
with \inc\ $\le 2$, or the 1400 galaxies without contamination (\cont\ = 0).

The dispersion in EMT-type per \bt\ attribute value seen in 
\fg\ref{emt_bt_hist} could simply be a binning effect becasue
there are 18 EM-types and 5 \bt\ attribute values, so that a spread
of $18/5\sim3.6$\ is expected for each \bt\ value. Moreover, the half 
confidence interval in the \bt\ attribute is 0.5 for 1504 galaxies,
1 for 2923 galaxies, and 1.5 for 98 EFIGI galaxies.
To evaluate whether there is some intrinsic dispersion in morphological type 
for a given \bt\ beyond the uncertainties in the visual eye
estimates of EMT and \bt, we performed a simple test.  
We assumed a linear relation between EM-type
and \bt\ with the two boundary conditions indicated in \fg\ref{emt_bt}:
\bt\ = 4 for EM-type = -5 (E), and \bt\ = 0 for EM-type = 10 (Im). 
The resulting relation is 
\bt$=-4/15\,\mathrm{EM-type} + 8/3$.\footnote{We disregard
   the \bt\ statistics for EM-type=-6 (cE), -4 (cD), and 11
  (dE) in this test, which contain fewer objects than the other classes.} 
For each value of EM-type in the catalogue, we then introduced
a dispersion in \bt\ by adding white noise in EM-type with a maximum
spread $N$, which projects onto a maximum spread of $4N/15$ in \bt.
We then rounded the values of \bt\ to the integer values 0 to 4,
and calculated the dispersion among these integer values for each
EM-type. This procedure has the advantage that it includes
the effect of discretisation of \bt.  The $\pm1\sigma$
dispersion in the measured \bt\ reaches values as high as $1.0$ for
most spiral types when $N=5$; for $N=6$, the dispersion reaches 
values of $1.2$ for several spiral types, which is too high. Note that $N=5$
is higher than the confidence intervals in the EM-types of 1, 2, and 3 
for every third of the EFIGI galaxies, and therefore cannot be the result solely
of these uncertainties in the estimated types.

In conclusion, a total spread of the EFIGI morphological types over five
types for a given \bt\ is necessary to reproduce the observed
dispersion in the discretised \bt\ for the various types. This is because
the Hubble sequence is primarily based on the
properties of the spiral arms such as their strength and curvature,
independently of the \bt\ ratio (see \sct\ref{hubseq}).


\subsection{Characteristics of the spiral arms              \label{arms}}
  
\begin{figure}
\resizebox{\hsize}{!}{\includegraphics{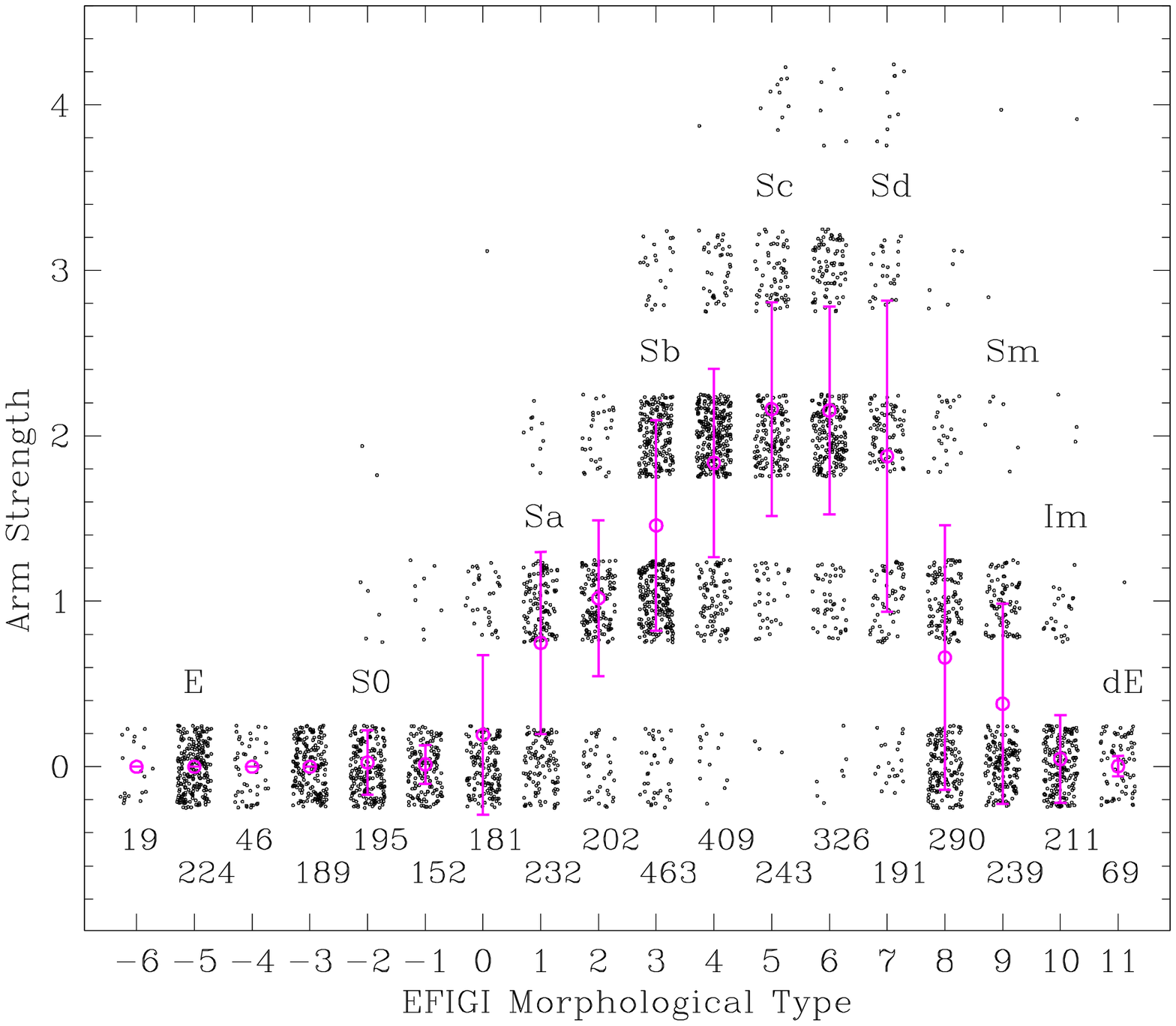}}
\resizebox{\hsize}{!}{\includegraphics{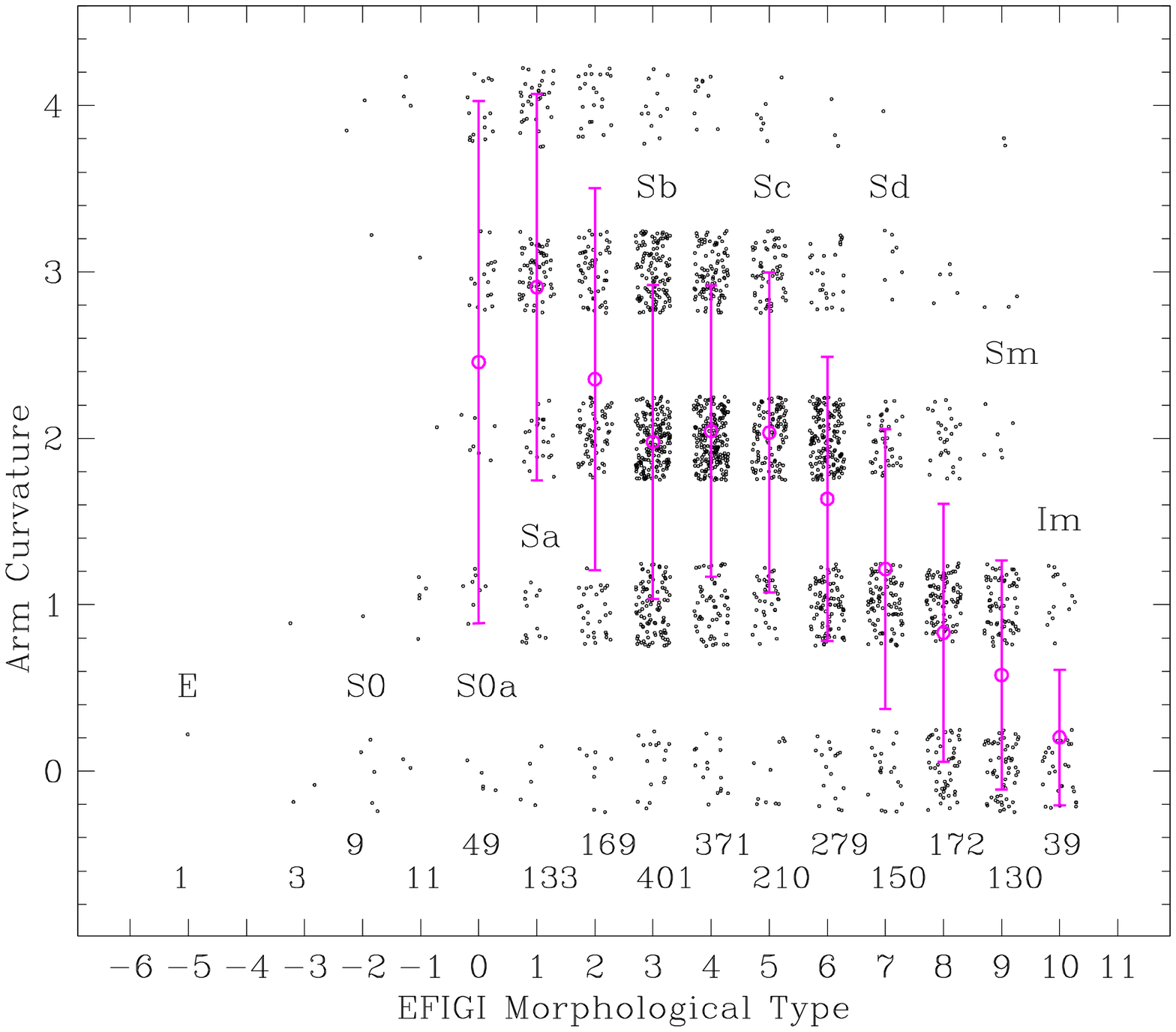}}
\caption{Distribution of the \arm\ (top) and \curv\
  attributes (bottom) for the 3881 and 2127 EFIGI galaxies for
  which these attributes are defined 
  (same presentation as in \fg\ref{emt_bt}). 
  The top panel shows the tight relation between
  the relative flux in the spiral arms and the morphological type,
  with an increase of the \arm\ from types Sa to Sc, then a
  decrease from types Sd to Im. The bottom panel shows a decrease of
  the \curv\ for increasing spiral types.}
\label{emt_arm}
\end{figure}

Two EFIGI attributes characterise the spiral arms: \arm, which
measures the fraction of the total flux included in the spiral arms,
and \curv, which measures the average curvature of the
spiral arms.

The top panel of \fg\ref{emt_arm} shows a strong correlation between \arm\
and EM-type. The graph is restricted to the 3881 galaxies 
for which the \arm\ attribute is defined. These galaxies also all have a 
confidence interval of 3 for this attribute.  
As expected, \arm\ is null for ellipticals, all lenticular types,
cE and cD galaxies. Then the fraction of flux in the spiral arm steeply increases
from types S0a up to Sc, and subsequently decreases from types Sd to Im. 
The \arm\ attribute measures the fraction of the total light of
the galaxy that is neither included in the bulge nor in the disk,
bar, or rings. Because the \bt\ ratio decreases with increasing EM-type,
the increase of \arm\ from types Sa to Sc indicates that the fraction
of light included in the spiral structure increases at the expense
of the bulge along these
types.  In contrast, the simultaneous decrease in the relative
fraction of light in the bulge and in the spiral arms from types Sd to
Im indicates an increasing fraction of light in the disk along these
types.

The diffuse spiral arms detected in the few S0 and S0$^+$ 
galaxies with \arm\ above 0 resemble the distorted ``plumes'' or ``loops'' of
matter seen in merging galaxies. Also, the few Sdm galaxies with
\arm\ = 3 or 4 have almost no disk emission, hence
although the spiral arms are very diffuse for those types, they
contain a major fraction of the galaxy flux in these particular
objects.

In the bottom panel of \fg\ref{emt_arm} one can observe a decrease
of \curv\ from spiral types S0a to Sm, where the Sa galaxies have the
most tightly wound arms and the Sm the most loosely wound arms of all
spirals.  Note, however, that the winding of the spiral arms shows a 
``plateau'' from Sb to Sc galaxies. Indeed, the Hubble type 
separation of these four types is based on an increasing continuity of
the spiral arm design, which is also based on a decrease in the patchiness of the 
dust, despite an increase in the arm flocculence (see \sct\ref{text}).


\subsection{Dynamical components and features: bars, rings,
  pseudo-rings and perturbation                       \label{dyn}}
  
Dynamical features of galaxies such as spiral arms, bars, and rings are
the major drivers of secular evolution via the flow of gas from the
disk to the central regions of galaxies (see the review by
\citealp{knapen10}). Mergers and interactions, which cause distorted
profiles, also contribute to this evolution. Bars allow a transfer of
angular momentum to the disk and the halo, and rings trace resonances in
the disk. 

\subsubsection{Bars                                   \label{bar}}
  
\begin{figure}
\centerline{\psfig{figure=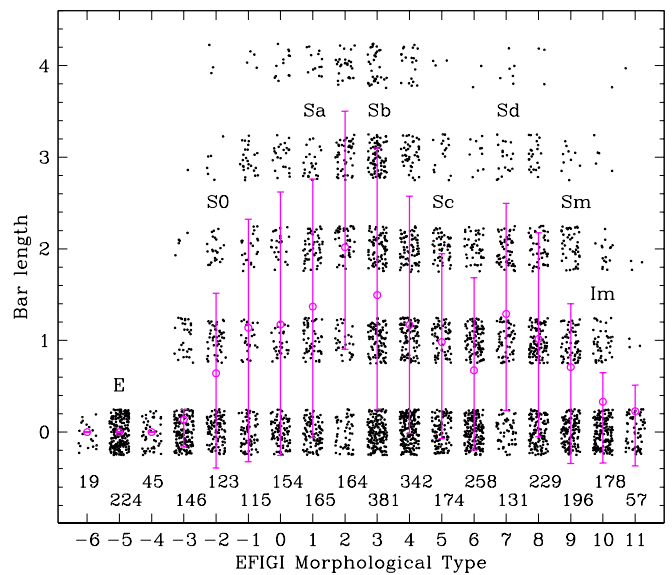,height=8cm,angle=0}}
\caption{Distribution of the \bbar\ attribute as a function of EFIGI
  morphological type for the 3878 EFIGI galaxies for which
  this attribute is defined (same presentation as in \fg\ref{emt_arm}). 
  This graph shows that bars are frequent among all galaxy types except 
  for the cE, E, cD, and dE galaxies, and the strongest bars lie in early-type spirals.}
\label{emt_bar}
\end{figure}
\begin{figure}
\centerline{\psfig{figure=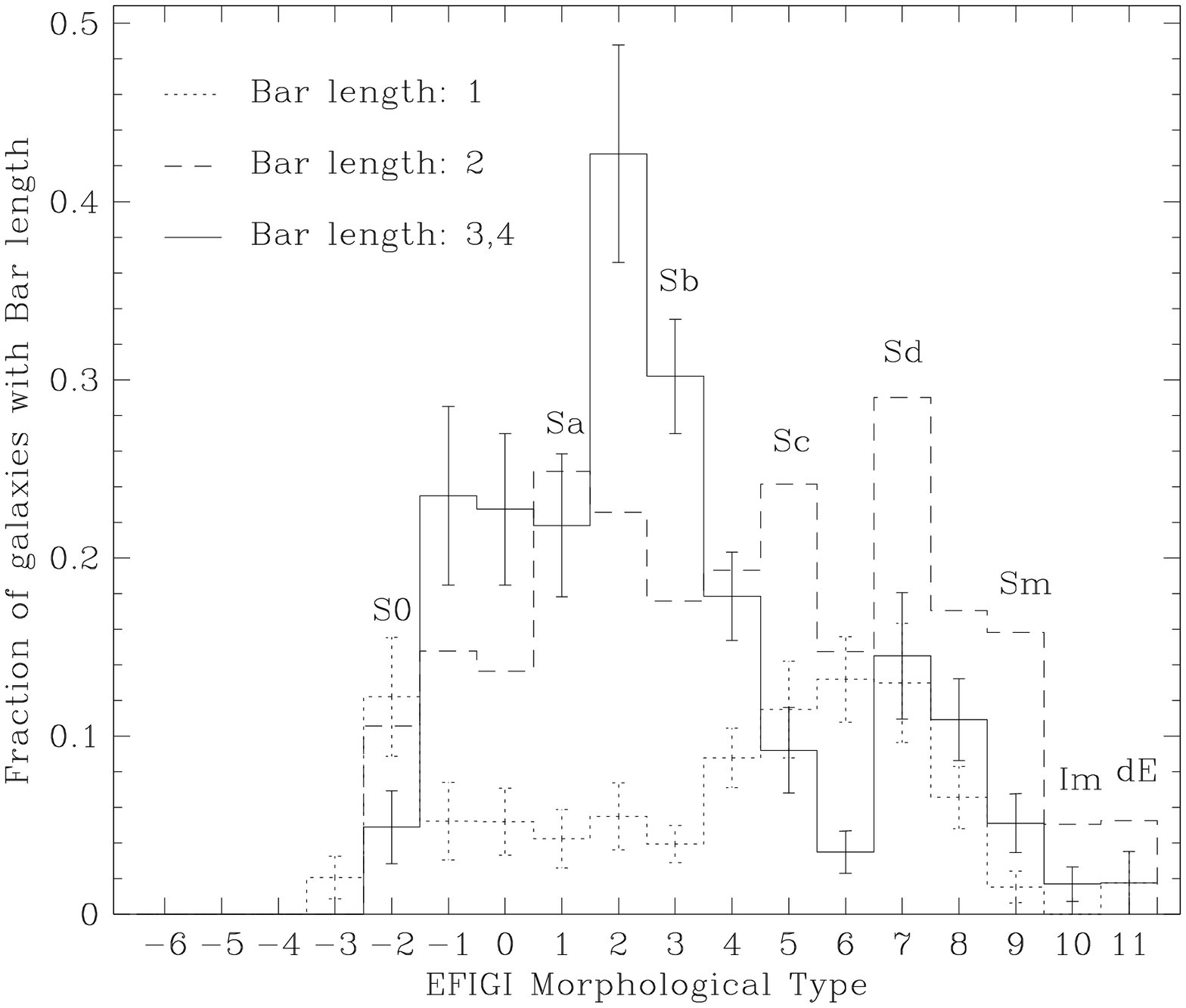,height=8cm,angle=0}}
\caption{Fraction of galaxies with different values of the 
  \bbar\ attribute as a function of EFIGI morphological type
  for the 3106 EFIGI galaxies with \incc\ $\le2$. This graph
  shows that bars are frequent among all galaxy types except E and dE,
  and the strongest bars lie in early-type spirals. For clarity, the Poisson
  error bars in the fractions with \bbar\ = 2 are not plotted; these are listed 
  in Table \ref{attrib_stat_bar} and are comparable to those for the other 
  plotted attribute values.}
\label{emt_bar_hist}
\end{figure}
\begin{table*}
  \caption{Statistics of the EFIGI \bbar\ attribute for \inc\ $\le2$, given as a percentage of galaxies with given attribute values}
\label{attrib_stat_bar}
\begin{center}
\begin{tabular}{lrrrrrrr}                      
\hline  
\hline
EFIGI Type       & \multicolumn{7}{c}{\tt Bar length}\\
                 &   \multicolumn{1}{c}{0}    &   \multicolumn{1}{c}{unsure} &    \multicolumn{1}{c}{1}    &     \multicolumn{1}{c}{2}    &    \multicolumn{1}{c}{3}    &     \multicolumn{1}{c}{4}  &  \multicolumn{1}{c}{1-2-3-4}     \\
\hline
cE cD E & 100$\pm$ 8 &       - &        - &        - &        - &        - &         - \\ 
S0$^-$&  87$\pm$11 & 11$\pm$ 3 &  2$\pm$ 1 &        - &        - &        - &   2$\pm$ 1 \\ 
  S0  &  59$\pm$ 9  & 14$\pm$ 4 & 12$\pm$ 3 & 11$\pm$ 3 &  2$\pm$ 1 &  2$\pm$ 1 &  28$\pm$ 5 \\ 
S0$^+$&  51$\pm$ 8  &  5$\pm$ 2 &  5$\pm$ 2 & 15$\pm$ 4 & 20$\pm$ 5 &  3$\pm$ 2 &  43$\pm$ 7 \\ 
 S0a  &  50$\pm$ 7  &  8$\pm$ 2 &  5$\pm$ 2 & 14$\pm$ 3 & 14$\pm$ 3 &  8$\pm$ 2 &  42$\pm$ 6 \\ 
  Sa  &  39$\pm$ 6  & 10$\pm$ 3 &  4$\pm$ 2 & 25$\pm$ 4 & 14$\pm$ 3 &  8$\pm$ 2 &  51$\pm$ 7 \\ 
 Sab  &  23$\pm$ 4  &  6$\pm$ 2 &  5$\pm$ 2 & 23$\pm$ 4 & 27$\pm$ 5 & 16$\pm$ 3 &  71$\pm$ 9 \\ 
  Sb  &  37$\pm$ 4  & 11$\pm$ 2 &  4$\pm$ 1 & 18$\pm$ 2 & 21$\pm$ 3 &  9$\pm$ 2 &  52$\pm$ 5 \\ 
 Sbc  &  44$\pm$ 4  & 11$\pm$ 2 &  9$\pm$ 2 & 19$\pm$ 3 & 12$\pm$ 2 &  6$\pm$ 1 &  46$\pm$ 4 \\ 
  Sc  &  45$\pm$ 6  & 10$\pm$ 2 & 11$\pm$ 3 & 24$\pm$ 4 &  9$\pm$ 2 &  1$\pm$ 1 &  45$\pm$ 6 \\ 
 Scd  &  55$\pm$ 6  & 14$\pm$ 2 & 13$\pm$ 2 & 15$\pm$ 3 &  3$\pm$ 1 &  1$\pm$ 1 &  31$\pm$ 4 \\ 
  Sd  &  32$\pm$ 6  & 11$\pm$ 3 & 13$\pm$ 3 & 29$\pm$ 5 & 11$\pm$ 3 &  3$\pm$ 2 &  56$\pm$ 8 \\ 
 Sdm  &  39$\pm$ 5  & 26$\pm$ 4 &  7$\pm$ 2 & 17$\pm$ 3 & 10$\pm$ 2 &  1$\pm$ 1 &  34$\pm$ 5 \\ 
  Sm  &  55$\pm$ 7  & 22$\pm$ 4 &  2$\pm$ 1 & 16$\pm$ 3 &  5$\pm$ 2 &       -  & 22$\pm$ 4 \\ 
  Im  &  77$\pm$ 9  & 16$\pm$ 3 &        - &  5$\pm$ 2 &  2$\pm$ 1 &       -  &  7$\pm$ 2 \\ 
  dE  &  88$\pm$17 &  4$\pm$ 3 &  2$\pm$ 2 &  5$\pm$ 3 &        - & 2$\pm$ 2  &  9$\pm$ 4 \\ 
\hline
S0$^-$ S0 S0$^+$& 67$\pm$ 5 & 10$\pm$ 2 &  6$\pm$ 1 &  8$\pm$ 1 &  7$\pm$ 1 &  2$\pm$ 1 & 23$\pm$ 3 \\ 
  Sa Sab        & 31$\pm$ 4 &  8$\pm$ 2 &  5$\pm$ 1 & 24$\pm$ 3 & 20$\pm$ 3 & 12$\pm$ 2 & 61$\pm$ 5 \\ 
  Sb Sbc        & 40$\pm$ 3 & 11$\pm$ 1 &  6$\pm$ 1 & 18$\pm$ 2 & 17$\pm$ 2 &  7$\pm$ 1 & 49$\pm$ 3 \\ 
  Sc Scd        & 51$\pm$ 4 & 12$\pm$ 2 & 12$\pm$ 2 & 18$\pm$ 2 &  5$\pm$ 1 &  1$\pm$ 1 & 37$\pm$ 3 \\ 
  Sdm Sm        & 47$\pm$ 4 & 24$\pm$ 3 &  4$\pm$ 1 & 16$\pm$ 2 &  7$\pm$ 1 &  1$\pm$ 1 & 29$\pm$ 3 \\ 
\hline
All disks       & 46$\pm$ 2 & 13$\pm$ 1 &  7$\pm$ 1 & 17$\pm$ 1 & 12$\pm$ 1 &  5$\pm$ 1 & 41$\pm$ 1 \\ 
All spirals     & 42$\pm$ 2 & 13$\pm$ 1 &  7$\pm$ 1 & 20$\pm$ 1 & 13$\pm$ 1 &  5$\pm$ 1 & 45$\pm$ 2 \\ 
All types       & 54$\pm$ 2 & 12$\pm$ 1 &  6$\pm$ 1 & 15$\pm$ 1 & 10$\pm$ 1 &  4$\pm$ 1 & 35$\pm$ 1 \\ 
\hline
\end{tabular}
\end{center}
{\it Notes:}
Null fractions are replaced by ``-'' for clarity.
``All disks'' corresponds to types from S0$^-$ to Sm, and ``All
      spirals'' to types from S0a to Sm.
\end{table*}

The EFIGI \bbar\ attribute measures the bar's length relative to the
galaxy $D_{25}$ isophotal diameter (see
\sct\ref{lim}). \fg\ref{emt_bar} shows the distribution of  \bbar\
for all EFIGI galaxies, except for the 580 galaxies for which
this attribute is undefined, which are essentially galaxies with
\inc\ = 3 or 4. Because the identification of bars is less reliable in highly
inclined disks, we restrict the analysis to the 3106 galaxies
with \inc\ $\le2$. We also separate as ``unsure'' the 348 and 10 galaxies
with \bbar\ = 1 and 2 respectively, which have a lower confidence limit set
to 0, because the probability that no bar exists at all in these galaxies 
was considered as non-negligible. To this end, special attention was brought to
setting the confidence limits for \bbar\ = 1. In the following, we also
use ``unsure'' bars as providers of more realistic error estimates.

\fg\ref{emt_bar_hist} shows the fractions of galaxies with \inc\ $\le2$ 
as a function of morphological type for the various values of the \bbar\ attribute.
The EFIGI cE, E, and cD galaxies have no detectable bars, whereas bars of
all lengths are visible in all other galaxy types, from lenticulars to
dE. The galaxy fractions are given in Table \ref{attrib_stat_bar} 
for each \bbar\ attribute value and for ``unsure'' bars; statistics for
groups of two or three types are also listed, as well as for ``all disks'', 
``all spirals'', and ``all types''.

Excluding galaxies with ``unsure'' bars, 183 ($6\pm1$\%) have a short 
bar, 462 ($15\pm1$\%) an intermediate bar, 310 ($10\pm1$\%) a strong bar, 
and 123 ($4\pm1$\%) a very strong bar; no bar was detected in 1660 galaxies 
($54\pm2$\%). Bars are therefore frequent in EFIGI galaxies, because they are present down 
to small sizes in $\sim35\pm1$\% of galaxies. The ``unsure'' bars were 
extracted from \bbar\ = 1 (see above), and a comparison of both columns 
of Table \ref{attrib_stat_bar} shows that they correspond to
more than half the total number of objects with an attribute value of 1.  
The separation of the galaxies with ``unsure'' 
bars from those with \bbar\ = 0 is also unclear, because 3/4 of the latter have
an upper confidence limit for a \bbar\ of 1, hence a ``short'' bar is not
excluded at the 70\% level. The total $12\pm1$\% of ``unsure'' bars 
therefore illustrate that the uncertainties in the frequency of 
galaxies with \bbar\ = 0 ($54\pm2$\%) or with \bbar\ = 1 ($6\pm1$\%) may be
significantly larger.

An examination of \fg\ref{emt_bar_hist} and Table \ref{attrib_stat_bar} 
shows that the long and very long bars (\bbar\ = 3 and 4) are preferentially
present in S0$+$, S0a,\ and early spiral types Sa to Sbc, with a peak for
Sab galaxies, which show a probability of $27\pm5$\% to have a long bar, and $16\pm3$\% a very
long bar. In contrast, Sm is the spiral galaxy type with the 
lowest fraction of bars altogether, which is also the case for long and very long bars,
whereas S0, Sc, Scd, and Sd types show the highest fraction of short bars (12-13\%). 
The Im and dE galaxies have a small but similar fraction of bars (7-9\%; bars
reside in the dS0 galaxies, which are included in the dE class). No very long bars are
detected in EFIGI Sm and Im types.

A comparison of the EFIGI bar frequencies yields results in rough agreement with
those stated by \citet{laurikainen09} for their selected sample of $\sim130$ 
S0  and $\sim200$ spiral galaxies, using the RC3 classes (see their Table 1). 
In particular, Table \ref{attrib_stat_bar} confirms that S0 galaxies have fewer 
and shorter bars than S0a and than all spirals together. 
Here, we add the detail that galaxies with the intermediate type S0$^-$ 
have only short bars, which are much less frequent than in S0 
(at least a factor 2, if one takes into account the ``unsure'' bars). 
In contrast, the rates of short, intermediate, and long bars in S0$^+$
galaxies are closer to those of S0a galaxies than in S0 galaxies.

For a more detailed comparison with Table 1 of \citet{laurikainen09}, we 
added in Table \ref{attrib_stat_bar} the fractions for the following
groups of galaxy types: S0$^-$-S0-S0$^+$, Sa-Sab, Sb-Sbc, Sc-Scd,
Sdm-Sm. In \citet{laurikainen09}, the fraction of bars of all lengths 
for the former four groups of types are $53\pm6\%$,
$65\pm7\%$, $69\pm6\%$, and $64\pm7\%$ respectively. Only the EFIGI fraction
for Sa-Sab ($61\pm6$\%) agrees within 1$\sigma$. For the other groups of types,
the EFIGI fractions are systematically lower, probably because of a poorer sensitivity, 
which has more impact for types with large fractions of short bars (S0 and 
late spirals). However, if one consider the EFIGI 8\% to 12\% of ``unsure'' bars 
as an estimate of the uncertainty, the differences between the type fractions
are reduced to 2$\sigma$. 

\begin{figure*}
\centerline{\psfig{figure=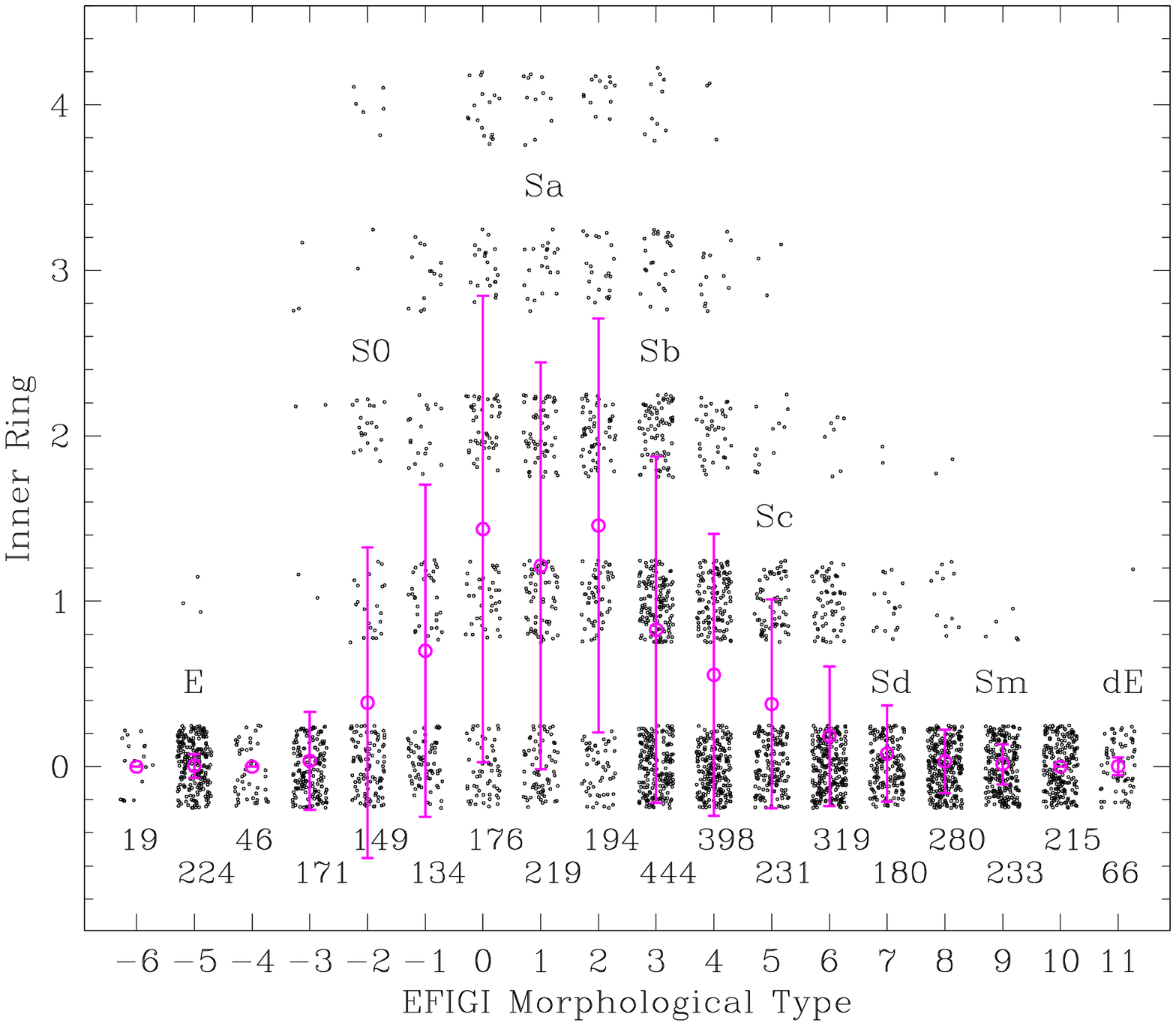,height=8cm,angle=0}
  \quad \psfig{figure=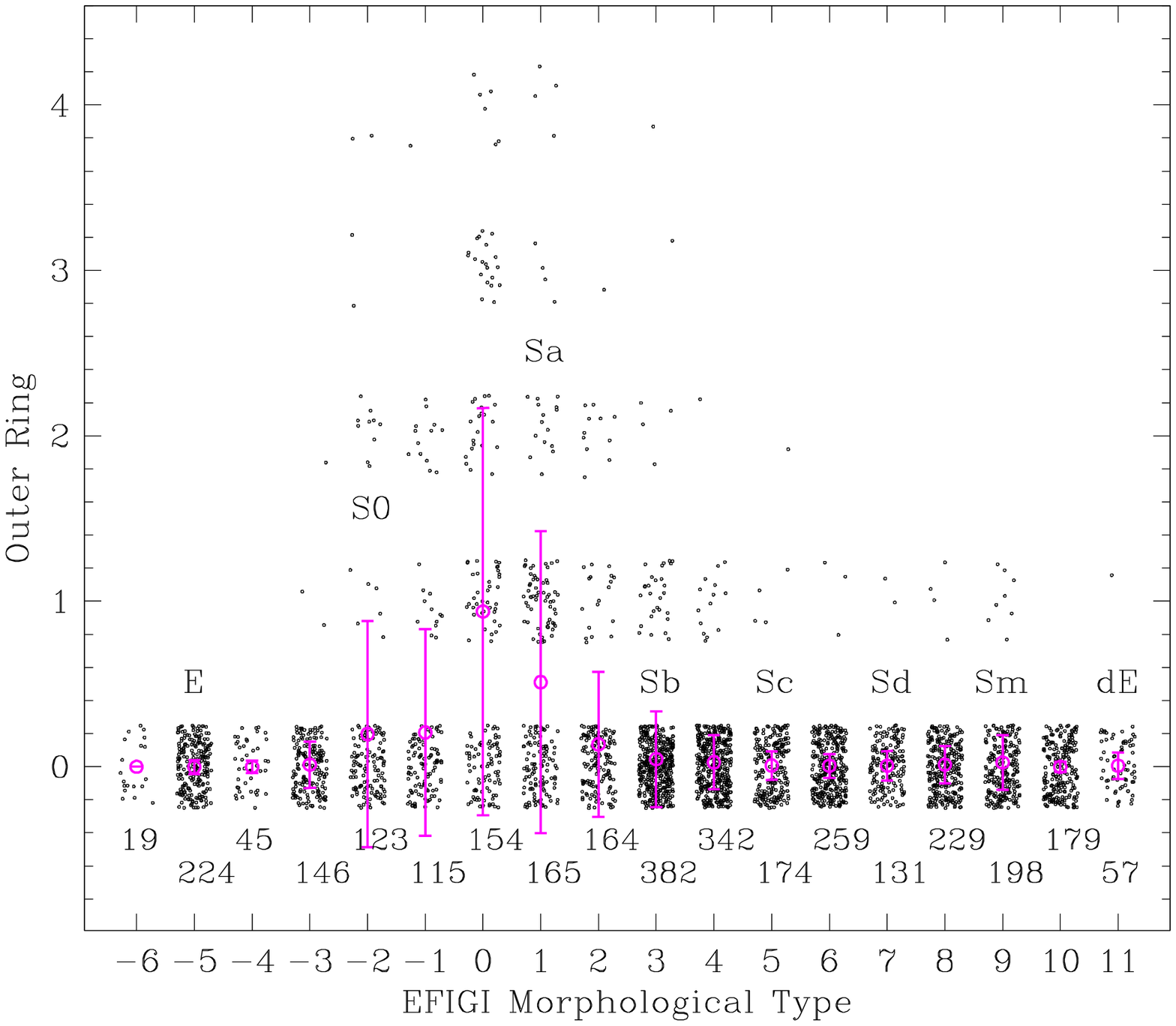,height=8cm,angle=0}}
\caption{Distribution of the \iring\ (left) and \oring\ (right)
attributes as a function of EFIGI morphological type for the 3698 and
3549 EFIGI galaxies for which these attributes are defined
(same presentation as in \fg\ref{emt_bt}). Strong inner rings dominate
among S0a, Sa, and Sab galaxies, whereas outer rings show a marked peak
in strength for S0a galaxies only.}  
\label{emt_ring}
\end{figure*}
\begin{figure*}
\centerline{\psfig{figure=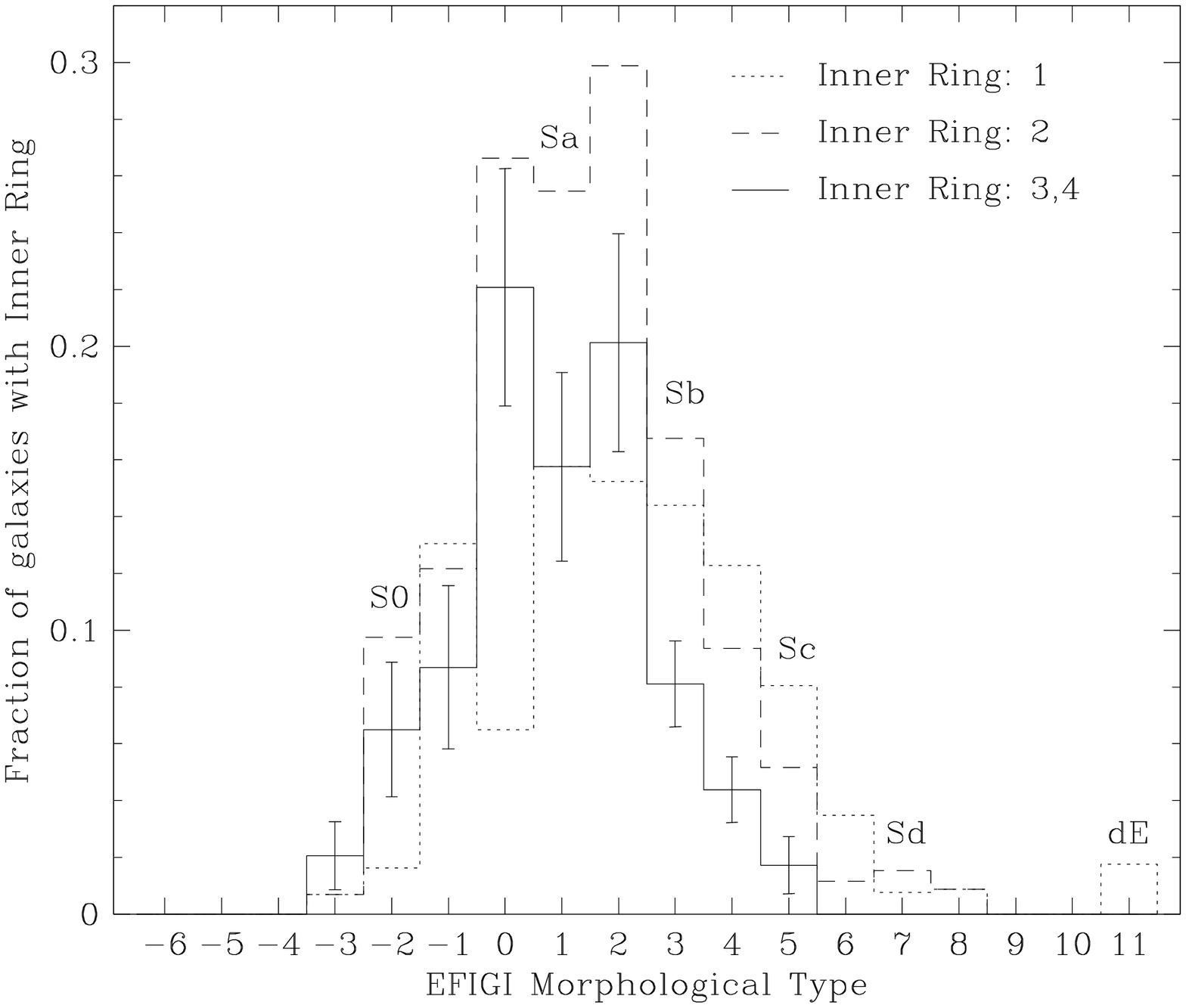,height=8cm,angle=0}
  \quad \psfig{figure=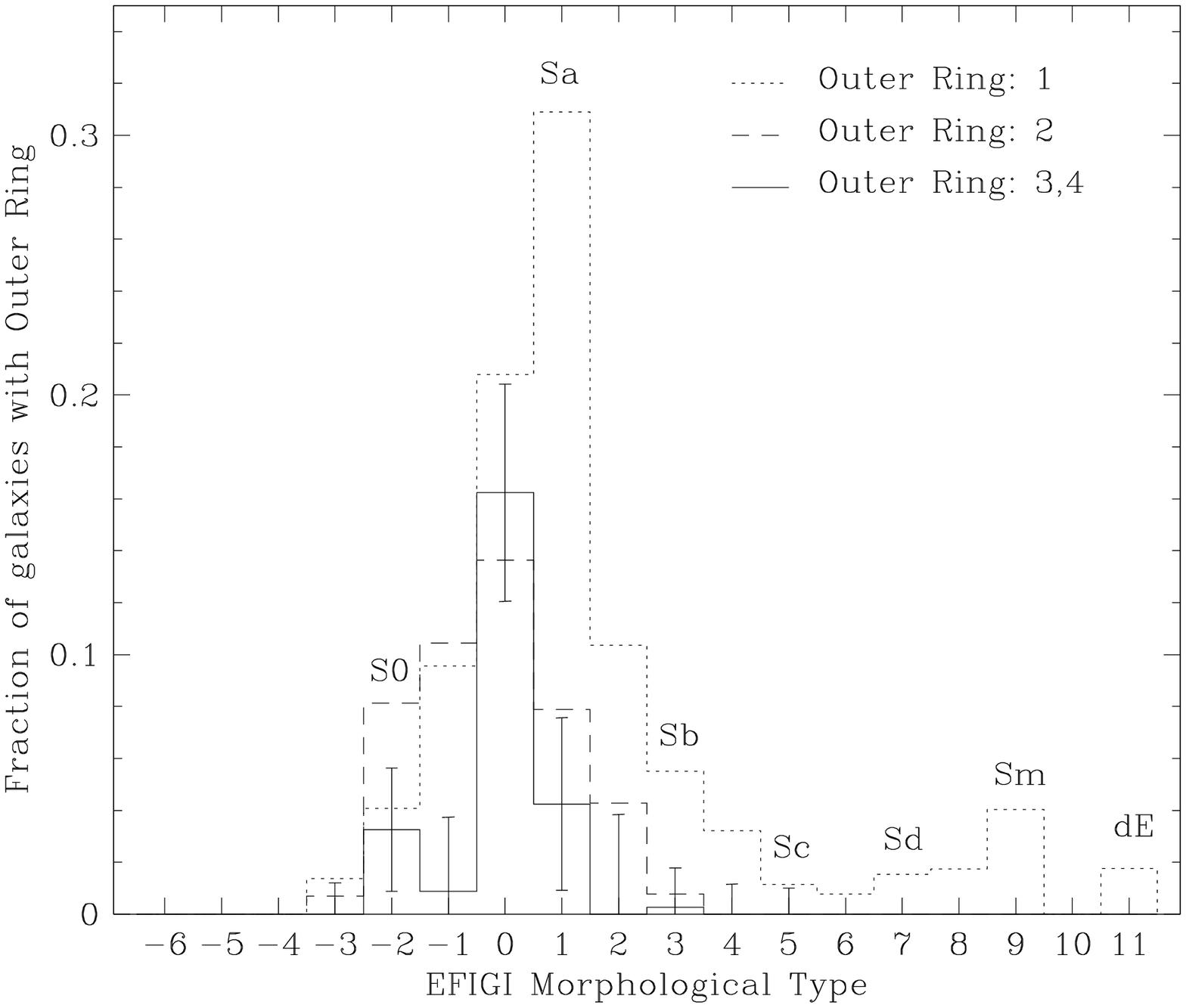,height=8cm,angle=0}}
\caption{Fraction of galaxies with different values of the \iring\ (left)
and \oring\ (right) attributes as a function of EFIGI morphological
type for the 3119 galaxies with \incc\ $\le2$. For clarity, only the Poisson
error bars in the fractions with \bbar\ = 3-4 are plotted.
This graph shows that the frequency for inner rings peaks
for S0a to Sab types, whereas the outer
rings are most frequent in S0a and Sa galaxies.}
\label{emt_ring_frac}
\end{figure*}

Although the four levels (``no bar'', ``short'', ``medium'', and ``long'') 
on which \citet{laurikainen09} grade bars complicate the comparison with 
the five levels of the EFIGI \bbar\ attribute, 
one can notice other differences between both samples. 
\citet{laurikainen09} quote S0a as the type with the highest fraction of bars 
($93\pm5\%$ compared to $42\pm6\%$ in the EFIGI catalogue), which is mostly owing 
to a higher fraction of strong bars than in the other types, whereas in
our sample the highest fraction of strong and very strong
bars (\bbar\ = 3 or 4) occurs in Sab. Moreover, the three largest fractions 
of bars in the EFIGI catalogue occur in Sab, Sb, and Sd, 
which show a probability of $71\pm9$\%, $52\pm5$\%, and $57\pm8$\% respectively 
to have a bar. This is because of $27\pm5$\% and 
$16\pm3$\% of galaxies with \bbar\ = 3 and 4 respectively in Sab, 
and $22\pm3$\% and $9\pm2$\% respectively in Sb, whereas the high frequency of bars 
in Sd is due to $13\pm3$\% and $29\pm5$\% of galaxies with \bbar\ = 1 and 2 respectively,
and only $12\pm3$\% and $3\pm2$\% with \bbar\ = 3 and 4 respectively.
These differences in bar fractions may be caused by the lack of spirals later than Scd 
and the reduced statistics in the $\sim330$ galaxy sample of \citet{laurikainen09} 
compared to the EFIGI catalogue; differences between the EFIGI \bbar\ attribute
definition and the RC3 bar classification may also be at play (see discussion in
\citealt{nair10}). 

\begin{table*}
  \caption{Statistics of the EFIGI \iring and \oring\ attributes
 for \inc\ $\le2$, given as a percentage of galaxies with given attribute values}
\label{attrib_stat_ring}
\begin{center}
\begin{tabular}{lrrrrrrrlrrrrrrr}                      
\hline  
\hline
EFIGI & \multicolumn{7}{c}{\tt Inner Ring}           & EFIGI & \multicolumn{6}{c}{\tt Outer Ring}      \\
 Type   &      0   &   unsure  &   1        &     2      &     3      &      4     & 1-2-3-4   &  Type   &   0        &   1       &   2        &  3         &  4         & 1-2-3-4  \\
\hline
cE cD E & 99$\pm$ 8 &        -  &         -  &         -  &         -  &         -  &         - & cE cD E &  98$\pm$ 8  &        -  &          - &        -   &        -   &        -  \\ 
S0$^-$&  96$\pm$11 & 1$\pm$ 1   &   1$\pm$ 1  &   1$\pm$ 1  &   2$\pm$ 1  &         -  &   3$\pm$ 2 & S0$^-$  & 97$\pm$11  &   1$\pm$ 1 &   1$\pm$ 1  &        -   &        -   &   2$\pm$ 1 \\ 
 S0   &  69$\pm$10 & 13$\pm$ 3  &   2$\pm$ 1  &  10$\pm$ 3  &   2$\pm$ 1  &   5$\pm$ 2  &  18$\pm$ 4 &  S0     &  85$\pm$11 &   4$\pm$ 2 &   8$\pm$ 3  &   2$\pm$ 1  &   2$\pm$ 1  &  15$\pm$ 4 \\ 
S0$^+$&  51$\pm$ 8  & 15$\pm$ 4  &  13$\pm$ 4  &  12$\pm$ 3  &   9$\pm$ 3  &         -  &  34$\pm$ 6 & S0$^+$  & 77$\pm$11  &  10$\pm$ 3 &  10$\pm$ 3  &        -   &   1$\pm$ 1  &  21$\pm$ 5 \\  
S0a   &  31$\pm$ 5  & 14$\pm$ 3  &   6$\pm$ 2  &  27$\pm$ 5  &  13$\pm$ 3  &   9$\pm$ 3  &  55$\pm$ 7 & S0a     &  49$\pm$ 7  &  21$\pm$ 4 &  14$\pm$ 3  &  13$\pm$ 3  &   3$\pm$ 1  &  51$\pm$ 7 \\  
 Sa   &  28$\pm$ 5  & 15$\pm$ 3  &  16$\pm$ 3  &  25$\pm$ 4  &  10$\pm$ 3  &   5$\pm$ 2  &  57$\pm$ 7 &  Sa     &  57$\pm$ 7  &  31$\pm$ 5 &   8$\pm$ 2  &   2$\pm$ 1  &   2$\pm$ 1  &  43$\pm$ 6 \\  
Sab   &  21$\pm$ 4  & 13$\pm$ 3  &  15$\pm$ 3  &  30$\pm$ 5  &  12$\pm$ 3  &   8$\pm$ 2  &  65$\pm$ 8 & Sab     &  84$\pm$10 &  10$\pm$ 3 &   4$\pm$ 2  &        -   &        -   &  15$\pm$ 3 \\ 
 Sb   &  42$\pm$ 4  & 18$\pm$ 2  &  14$\pm$ 2  &  17$\pm$ 2  &   5$\pm$ 1  &   3$\pm$ 1  &  39$\pm$ 4 &  Sb     &  93$\pm$ 7  &   5$\pm$ 1 &   1$\pm$ 1  &        -   &   0$\pm$ 1  &   7$\pm$ 1 \\ 
Sbc   &  53$\pm$ 5  & 20$\pm$ 3  &  12$\pm$ 2  &   9$\pm$ 2  &   4$\pm$ 1  &   1$\pm$ 1  &  26$\pm$ 3 & Sbc     &  96$\pm$ 7  &   3$\pm$ 1 &          - &        -   &        -   &   3$\pm$ 1 \\ 
 Sc   &  58$\pm$ 7  & 27$\pm$ 4  &   8$\pm$ 2  &   5$\pm$ 2  &   2$\pm$ 1  &         -  &  15$\pm$ 3 &  Sc     &  99$\pm$11 &   1$\pm$ 1 &        -   &        -   &        -   &   1$\pm$ 1 \\  
Scd   &  77$\pm$ 7  & 19$\pm$ 3  &   3$\pm$ 1  &   1$\pm$ 1  &         -  &         -  &   5$\pm$ 1 & Scd     &  99$\pm$ 9  &   1$\pm$ 1 &        -   &        -   &        -   &   1$\pm$ 1 \\ 
 Sd   &  87$\pm$11 & 11$\pm$ 3  &   1$\pm$ 1  &   2$\pm$ 1  &         -  &         -  &   2$\pm$ 1 &  Sd     &  98$\pm$12 &   2$\pm$ 1 &        -   &        -   &        -   &   2$\pm$ 1 \\ 
Sdm   &  95$\pm$ 9  &  3$\pm$ 1  &   1$\pm$ 1  &   1$\pm$ 1  &         -  &         -  &   2$\pm$ 1 & Sdm     &  98$\pm$ 9  &   2$\pm$ 1 &        -   &        -   &        -   &   2$\pm$ 1 \\ 
 Sm   &  97$\pm$10 &  3$\pm$ 1  &         -  &         -  &         -  &         -  &         - &  Sm     &  95$\pm$10 &   4$\pm$ 1 &        -   &        -   &        -   &   4$\pm$ 1 \\ 
 Im   & 100$\pm$11 &        -  &         -  &         -  &         -  &         -  &         - &  Im     &  99$\pm$10 &        -  &        -   &        -   &        -   &        -  \\ 
 dE   &  98$\pm$18 &        -  &   2$\pm$ 2  &         -  &         -  &         -  &   2$\pm$ 2 &  dE     &  98$\pm$18 &   2$\pm$ 2 &        -   &        -   &        -   &   2$\pm$ 2 \\ 
\hline
All disks &  61$\pm$ 2 & 14$\pm$ 1 &  8$\pm$ 1 &  10$\pm$ 1  &   4$\pm$ 1  &   2$\pm$ 1  &  25$\pm$ 1  &  All disks &  89$\pm$ 2 &  7$\pm$ 1 &   3$\pm$ 1  &   1$\pm$ 1  &   1$\pm$ 1  &  11$\pm$ 1  \\ 
All types &  68$\pm$ 2 & 12$\pm$ 1 &  7$\pm$ 1 &   9$\pm$ 1  &   4$\pm$ 1  &   2$\pm$ 1  &  21$\pm$ 1  &  All types &  91$\pm$ 2 &  5$\pm$ 1 &   2$\pm$ 1  &   1$\pm$ 1  &   0$\pm$ 1  &   9$\pm$ 1  \\ 
\hline
\end{tabular}
\end{center}
{\it Notes:}
Null fractions are replaced by `` - '' for clarity.
``All disks'' corresponds to types from S0$^-$ to Sm.
\end{table*}

The EFIGI statistics of the \bbar\ attribute differ from the minimum in bar
fraction at stage Sc and the very high bar fraction at stage Sm found 
in \citet[ see their \fg 1.14]{buta07} for a subsample in isophotal 
diameter and isophotal elongation.
The EFIGI results also do not confirm the increasing fraction of 
bars in disk-dominated, hence late-type spirals, obtained by 
\citet{barazza08}, using ellipse-fitting of bars in SDSS $r$-band 
images of $\sim 3700$ disk galaxies: the last two columns of Table 
\ref{attrib_stat_bar} show that the EFIGI fraction of bars of all lengths
peaks for Sab galaxies, and regularly decreases for later types. 
These authors however find a $\sim50$\% overall percentage of bars in disk galaxies,
which agrees well with the 50-60\% value
obtained by \citet{knapen00} using the RC3 bar classification. 
Yet the 12\% of EFIGI ``unsure'' bars for ``All types'' reduces 
the EFIGI 35\% overall rate of bars to a 1$\sigma$ 
difference compared with those measured 
by \citet{barazza08}, \citet{laurikainen09}, and \citet{knapen00}. 
\citet{nair10} find an even smaller fraction of 25\%
of bars in $\sim 14000$ SDSS galaxies, despite a consistent scale of the bar 
``strength'' with EFIGI (Paper I). This illustrates the impact of 
the different selection criteria (threshold, intensity scale etc.), 
even when using the same image material.

\subsubsection{Rings                                  \label{rings}}

The strength of the three EFIGI ring attributes (\iring, \oring, and
\pseu) is a monotonically increasing function of the fraction of the
galaxy flux (dominated by the $g$-band), which is held in each feature
relative to that of the whole galaxy; each strength scale also depends
on the most extreme cases found in the EFIGI catalogue.

\fg\ref{emt_ring} shows the distribution of the \iring\ and \oring\
attributes for all EFIGI galaxies: both graphs show systematic variations
in ring strength, with peaks for early-type spirals.
For further examination, we list in Table \ref{attrib_stat_ring} and plot
in \fg\ref{emt_ring_frac} the various statistics of the \iring\ and \oring\ 
attributes.

We again restricted the analysis to the 3106 galaxies 
with \inc\ $\le2$, because ring attributes are difficult to determine for
highly inclined disks. Indeed, \iring\ and \oring\ are defined 
for only 6\% to 4\% galaxies with \inc\ = 4. Although these proportions
increase to 80-63\% for \inc\ = 3, the fractions of galaxies for a given
attribute value of \iring\ and \oring\ show a poor correlation with
the fractions for \inc\ $\le2$. For \iring, we also separated as ``unsure'' 
rings the 360 and 4 galaxies with \iring\ = 1 and 2 respectively, which have a 
lower confidence limit set to 0; we were not able to perform
this distinction during the homogenisation process of the \oring\ and \pseu\ 
attributes, owing to the very low contrast of both types of feature.

The left panel of \fg\ref{emt_ring_frac} shows that inner rings are most 
frequent in S0a, Sa, and Sab types, occurring in $55\pm7$\%, $57\pm7$\% and 
$65\pm8$\% of these objects (see Table \ref{attrib_stat_ring}). 
Together with S0$^+$, these EM-types also have the highest fractions of strong 
and/or very strong inner rings. Interestingly, weak and moderate inner rings 
also show a peak in frequency for Sa and Sab respectively. Altogether,
inner rings are present in $23\pm1$\% of disks (types S0$^-$ to Sd), and in 
$21\pm1$\% of galaxies of all types. For all attribute values, the frequency 
of objects decreases for types earlier than S0a, and for types later than Sab. 
Finally, S0$^-$, Scd, Sd, Sm, and dE galaxies have \iring\ fractions between
2 and 5\%, whereas E, cE, cD, Sm, and Im have no internal rings at all. 

Outer rings are less frequent than inner rings by nearly a factor of 2 
and occur in only $11\pm1$\% of disk galaxies, and $9\pm1$\% of all EFIGI 
galaxies, as shown in Table \ref{attrib_stat_ring}. 
Moreover, the distribution of outer rings is 
shifted towards earlier types, with a peak for S0a galaxies (right panel of 
\fg\ref{emt_ring_frac}), as expected from the definition of this EM-type. 
Earlier S0-S0$^+$ types on the one hand, as well as later Sa-Sab types
have frequencies of outer rings between $15\pm3$\% to $43\pm3$\%.
Outer rings occur in less than 5\% of S0$^-$, Sbc to Sm, and dE types,
and are completely absent from E, cE, cD, and Im galaxies. 

\fg\ref{emt_pseu_frac} shows the frequency distribution of the
small population of galaxies (less than 100) that have outer pseudo-rings
including the $R_1^\prime$ ``eight-shape'' pattern, and the $R_2^\prime$ and intermediate
$R_1R_2^\prime$ patterns, as defined by \citet[][ see Paper I for details of 
this attribute strength and for colour images of pseudo-rings; see also \citealp{buta95} 
for blue images]{buta96}.
These features occur essentially in S0$^+$, S0a, Sa, Sab, and Sb galaxies,
with a marked peak for Sab ($13\pm3$\%), and frequencies of 3 to 8\%
for the other types. \fg\ref{emt_pseu_frac} also shows that
although \pseu\ = 1 and 2 peak for Sab, the frequency of galaxies with 
\pseu\ = 3 and 4 regularly decreases from S0a to Sbc.
The lower fraction of outer rings compared to inner rings and the 
even lower fraction of
outer pseudo-rings was also found in the RC3 catalogue by \citet{buta96}.
All these objects are a sub-population of galaxies with a bar, 
as shown below.

\begin{table}
  \caption{Statistics of the EFIGI \pseu\ attribute
 for \inc\ $\le2$ given as a percentage of galaxies with given attribute values}
\label{attrib_stat_pseudo}
\begin{center}
\begin{tabular}{lrrrrrrr}                      
\hline  
\hline
EFIGI & \multicolumn{6}{c}{\tt Pseudo-Ring}\\
 Type     &   0    &   1    &   2    &  3      &  4     & 1-2-3-4  \\
\hline
S0$^+$ &  95$\pm$13 &   3$\pm$ 2 &         -  &         -  &         -  &   3$\pm$ 2 \\ 
S0a  &   92$\pm$11  &   3$\pm$ 1 &   2$\pm$ 1  &   2$\pm$ 1  &   1$\pm$ 1  &   7$\pm$ 2 \\ 
 Sa  &   92$\pm$10  &   5$\pm$ 2 &   1$\pm$ 1  &   1$\pm$ 1  &   1$\pm$ 1  &   8$\pm$ 2 \\ 
Sab  &   85$\pm$10  &   8$\pm$ 2 &   4$\pm$ 2  &   1$\pm$ 1  &   1$\pm$ 1  &  13$\pm$ 3 \\ 
 Sb  &   93$\pm$ 7   &   3$\pm$ 1 &   2$\pm$ 1  &   1$\pm$ 1  &   1$\pm$ 1  &   6$\pm$ 1 \\ 
Sbc  &   99$\pm$ 8   &   0$\pm$ 1 &         -  &        -   &   0$\pm$ 1  &   1$\pm$ 1 \\ 
\hline 
 S0$^+$ to Sbc & 95$\pm$ 3 & 3$\pm$ 1 & 1$\pm$ 1  &   1$\pm$ 1  &   1$\pm$ 1  &   5$\pm$ 1  \\ 
All types      & 97$\pm$ 2 & 1$\pm$ 1 & 1$\pm$ 1  &   0$\pm$ 1  &   0$\pm$ 1  &   2$\pm$ 1  \\ 
\hline 
\end{tabular}
\end{center}
{\it Notes:}
Null fractions are replaced by `` - '' for clarity.
``All disks'' corresponds to types from S0$^-$ to Sm.
Unlisted galaxy types have \pseu\ = 0.
\end{table}

\begin{figure}
\resizebox{\hsize}{!}{\includegraphics{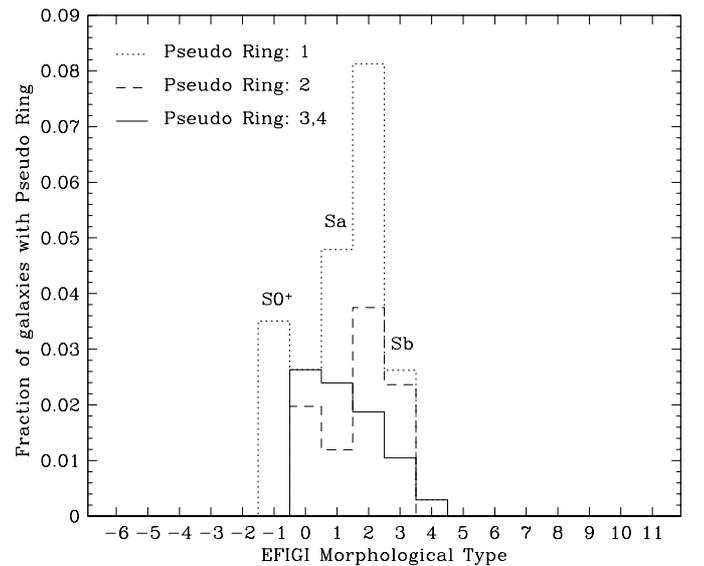}}
\caption{Fraction of galaxies with different values of the 
\pseu\ attribute as a function of EFIGI morphological type for the 3105
galaxies with \incc\ $\le2$. For clarity, no errorbars are plotted,
as these are large due to the small number of objects (see Table 
\ref{attrib_stat_pseudo}). The outer pseudo-ring pattern is essentially
present in S0$^+$, S0a, Sa, Sab and Sb galaxies.}  
\label{emt_pseu_frac}
\end{figure}

We list in Table \ref{attrib_cross} the fractions of galaxies
with \bbar, \iring, \oring, and \pseu\ attributes equal to 0 or in the
1-4 interval (listed in the Table header), which have one of the other 
attribute in the 1-4 interval (``unsure'' galaxies with \bbar\ = 1-2 or 
\iring\ = 1-2 are discarded).
First, this Table shows that the percentage of barred galaxies is similar
among galaxies with inner or outer rings ($66\pm4$\% and $54\pm5$\% respectively). 
However, inner rings are twice as frequent in barred galaxies
($39\pm2$\%) than in non-barred galaxies ($11\pm1$\%), and outer rings are 3.5 times
more frequent ($10\pm1$\% and $6\pm1$\% of barred and non-barred galaxies respectively).
Moreover, outer rings are two to four times less frequent than inner rings in 
non-barred and barred galaxies.

Both types of rings also appear to be tightly correlated: inner rings are four times 
more frequent in galaxies with an outer ring ($64\pm6$\%) than without ($16\pm1$\%); 
outer rings are nine times more frequent in galaxies with an inner ring ($27\pm2$\%)
than without ($3\pm1$\%). These statistics also indicate that an outer ring 
implies the presence of an inner ring with a $64\pm6$\% probability, whereas 
an inner ring implies an outer ring with a $27\pm2$\% probability.

Table \ref{attrib_cross} confirms quantitatively that the vast majority
of pseudo-rings also contain a bar ($95\pm15$\%) and an inner ring ($84\pm14$\%),
these structural properties being most likely dynamically related. In comparison,
only $33\pm1$\% of galaxies with no pseudo-ring host 
a bar and $19\pm1$\% an inner ring. Pseudo-rings are indeed nearly 50 times more
frequent in barred ($11\pm1$\%) than in non-barred galaxies ($0.2\pm0.2$\%), 
and about 10 times more frequent in galaxies with an inner ring ($12\pm2$\%) than without
($0.9\pm0.4$\%). Note also that $58\pm11$\% of galaxies with a pseudo-ring are 
classified as having an outer ring: this occurs when the pseudo-ring
is sufficiently round to mimic a single ring; as a result, pseudo-rings are six times 
more frequent in galaxies with an outer ring ($19\pm3$\%) than without ($3\pm1$\%).

\begin{table*}
  \caption{Correlations of the EFIGI \bbar, \iring, \oring, and \pseu\
 attributes for \inc\ $\le2$, measured as the percentage of galaxies
 among the intervals of attributes listed in the header that have
 another attribute in the interval listed in the left column}  
\label{attrib_cross}
\begin{center}
\begin{tabular}{lrrrrrrrr}                   
\hline  
\hline
           & \multicolumn{2}{c}{\bbar}  & \multicolumn{2}{c}{\iring} 
           & \multicolumn{2}{c}{\oring} & \multicolumn{2}{c}{\pseu} \\
           &   0  &  1-4 &  0   &  1-4 &  0   &  1-4 &  0   &  1-4 \\
\hline   
\bbar\ = 1-4  &   -        &  -      & $24\pm1$   & $66\pm4$& $33\pm1$& $54\pm5$& $33\pm1$& $95\pm15$\\ 
\iring\ = 1-4 & $11\pm1$   & $39\pm2$&  -         &   -     & $16\pm1$& $64\pm6$& $19\pm1$& $84\pm14$\\ 
\oring\ = 1-4 & $6\pm1$    & $10\pm1$&  $3\pm1$   & $27\pm2$&  -      &  -      & $8\pm1$ & $58\pm11$\\ 
\pseu\ = 1-4  & $0.2\pm0.2$& $11\pm1$& $0.9\pm0.4$& $12\pm2$& $3\pm1$ & $19\pm3$&   -     &  -       \\
\hline
\end{tabular}
\end{center}
{\it Notes:}
For clarity, we indicate as `` - '' the 0\% or 100\% fractions
       between one attribute and itself.
For \pseu=1-4, only galaxy types from S0$^+$ to Sbc that
  host pseudo-rings (see Table \ref{attrib_stat_ring}) are considered; fractions smaller 
  than 1\% are given to a tenth of a percent.
\end{table*}

\subsubsection{Perturbation                           \label{pert}}               

\begin{figure}
\resizebox{\hsize}{!}{\includegraphics{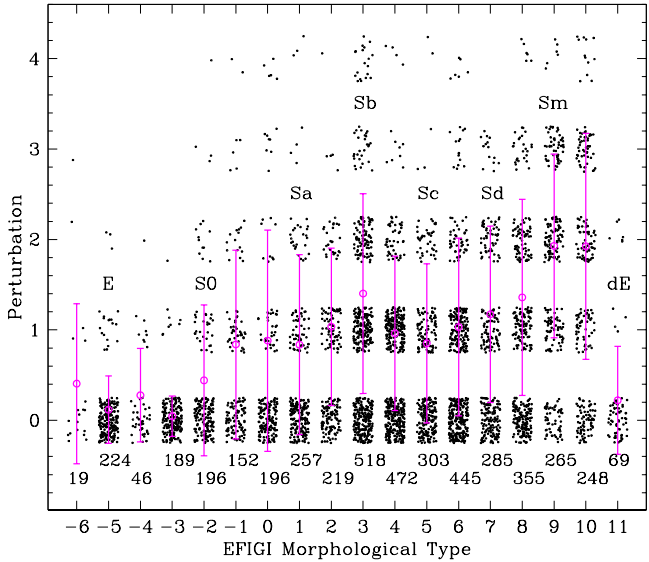}}
\caption{Distribution of the \pert\ attribute for all 4458 EFIGI
  galaxies (same presentation as in \fg\ref{emt_bt}). This graph shows
  the increasing \pert\ for types Sd, Sdm, Sm, and Im.}
\label{emt_pert}
\end{figure}
\begin{table}
  \caption{Statistics of the EFIGI \pert\ attribute given as a percentage of galaxies with given attribute values}
\label{attrib_stat_pert}
\begin{center}
\begin{tabular}{lrrrrrr}                   
\hline  
\hline
EFIGI  & \multicolumn{6}{c}{\tt Perturbation}\\
 Type  &    0      &    1      &    2      &     3     &    4     & 1-2-3-4  \\
\hline
  cE   & 74$\pm$26 & 16$\pm$10 &  5$\pm$ 5 & 5$\pm$ 5  &     -    & 26$\pm$13\\ 
  cD   & 80$\pm$18 & 17$\pm$ 7 &  2$\pm$ 2 &     -     &     -    & 20$\pm$ 7\\ 
   E   & 91$\pm$ 9 &  8$\pm$ 2 &  1$\pm$ 1 &     -     &     -    &  9$\pm$ 2\\ 
S0$^-$ & 95$\pm$10 &  4$\pm$ 2 &  1$\pm$ 1 &     -     &     -    &  5$\pm$ 2\\ 
  S0   & 78$\pm$ 8 & 14$\pm$ 3 &  6$\pm$ 2 &  2$\pm$ 1 & 1$\pm$ 1 & 22$\pm$ 4\\ 
S0$^+$ & 60$\pm$ 8 & 26$\pm$ 5 & 10$\pm$ 3 &  3$\pm$ 1 & 1$\pm$ 1 & 40$\pm$ 6\\ 
 S0a   & 72$\pm$ 8 & 16$\pm$ 3 &  5$\pm$ 2 &  4$\pm$ 1 & 3$\pm$ 1 & 28$\pm$ 4\\ 
  Sa   & 67$\pm$ 7 & 21$\pm$ 3 & 10$\pm$ 2 &  2$\pm$ 1 & 1$\pm$ 1 & 33$\pm$ 4\\ 
 Sab   & 52$\pm$ 6 & 33$\pm$ 5 & 13$\pm$ 3 &  2$\pm$ 1 & 0$\pm$ 1 & 48$\pm$ 6\\ 
  Sb   & 40$\pm$ 3 & 33$\pm$ 3 & 18$\pm$ 2 &  5$\pm$ 1 & 4$\pm$ 1 & 60$\pm$ 4\\ 
 Sbc   & 49$\pm$ 4 & 38$\pm$ 3 & 10$\pm$ 2 &  2$\pm$ 1 & 1$\pm$ 1 & 51$\pm$ 4\\ 
  Sc   & 56$\pm$ 5 & 30$\pm$ 4 & 13$\pm$ 2 &  1$\pm$ 1 & 1$\pm$ 1 & 44$\pm$ 5\\ 
 Scd   & 51$\pm$ 4 & 34$\pm$ 3 & 10$\pm$ 2 &  3$\pm$ 1 & 1$\pm$ 1 & 49$\pm$ 4\\ 
  Sd   & 49$\pm$ 5 & 28$\pm$ 4 & 18$\pm$ 3 &  5$\pm$ 1 &     -    & 51$\pm$ 5\\ 
 Sdm   & 42$\pm$ 4 & 29$\pm$ 3 & 20$\pm$ 3 &  7$\pm$ 1 & 2$\pm$ 1 & 58$\pm$ 5\\ 
  Sm   & 21$\pm$ 3 & 25$\pm$ 3 & 32$\pm$ 4 & 21$\pm$ 3 & 2$\pm$ 1 & 79$\pm$ 7\\ 
  Im   & 29$\pm$ 4 & 21$\pm$ 3 & 25$\pm$ 4 & 20$\pm$ 3 & 5$\pm$ 1 & 71$\pm$ 7\\ 
  dE   & 86$\pm$15 &  6$\pm$ 3 &  9$\pm$ 4 &     -     &     -    & 14$\pm$ 5\\ 
\hline
All spirals & 47$\pm$ 1 & 31$\pm$ 1 & 15$\pm$ 1 & 5$\pm$ 1 & 2$\pm$ 1 & 53$\pm$ 2 \\ 
All types   & 54$\pm$ 1 & 26$\pm$ 1 & 13$\pm$ 1 & 5$\pm$ 1 & 2$\pm$ 1 & 46$\pm$ 1 \\ 
\hline
\end{tabular}
\end{center}
{\it Note:} Null fractions are replaced by `` - '' for clarity.
\end{table}

The deviations from smooth and/or symmetric profiles have been 
quantified by \citet{conselice06}. These features are also of a dynamical 
origin, because numerical simulations show that morphological perturbation 
is the sign of interactions between galaxies \citep{teyssier10}.

\fg\ref{emt_pert} shows that the EFIGI
\pert\ attribute increases for the latest spiral types Sd, Sdm,
and Sm ; for Im types, the attribute is as high as for Sm
galaxies. Table \ref{attrib_stat_pert} shows that between 1/3 to 2/3 of
types from S0$^+$ to Sdm have perturbed profiles, and the percentage increases 
to $79\pm7$\% for Sm and $71\pm7$\% for Im galaxies. Altogether, $53\pm2$\% 
of the spiral galaxies exhibit some distortion in their profile.
In contrast, only $22\pm4$\% of S0, $9\pm2$\% of E and $5\pm2$\% of S0$^-$ 
are perturbed. 

Table \ref{attrib_stat_pert} also shows that Sm and Im types are more
frequently strongly perturbed (\pert\ = 3) than earlier spiral types: 
this attribute value occurs in $20$ to $21\pm3$\% of Sm-Im galaxies and in less 
than $7\pm1$\% of earlier spiral types and S0. One possible origin for
the higher rate of detectable perturbed profiles in Sm and Im galaxies 
could be their lower luminosity (de Lapparent \& Bertin 2011b, in prep{.}), hence mass, which 
might make them more sensitive to tidal effects by neighbouring galaxies.
That Im galaxies are more numerous in the dense environments of
clusters of galaxies might contribute to increasing their frequency
of tidal distortions.


\subsection{Texture: dust, flocculence, hot spots   \label{text}}
  
\begin{figure*}
\centerline{\psfig{figure=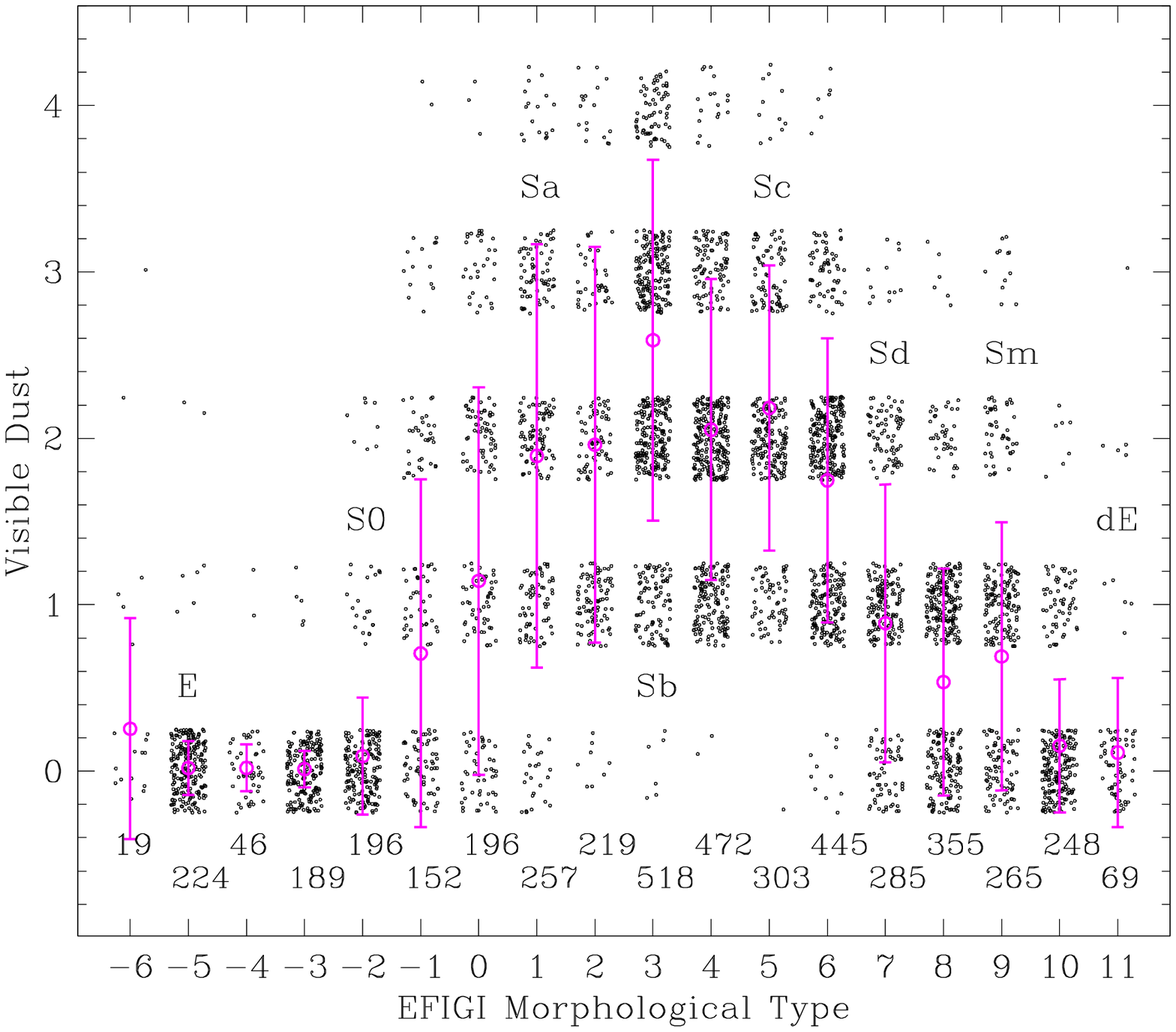,height=8cm,angle=0}
  \quad \psfig{figure=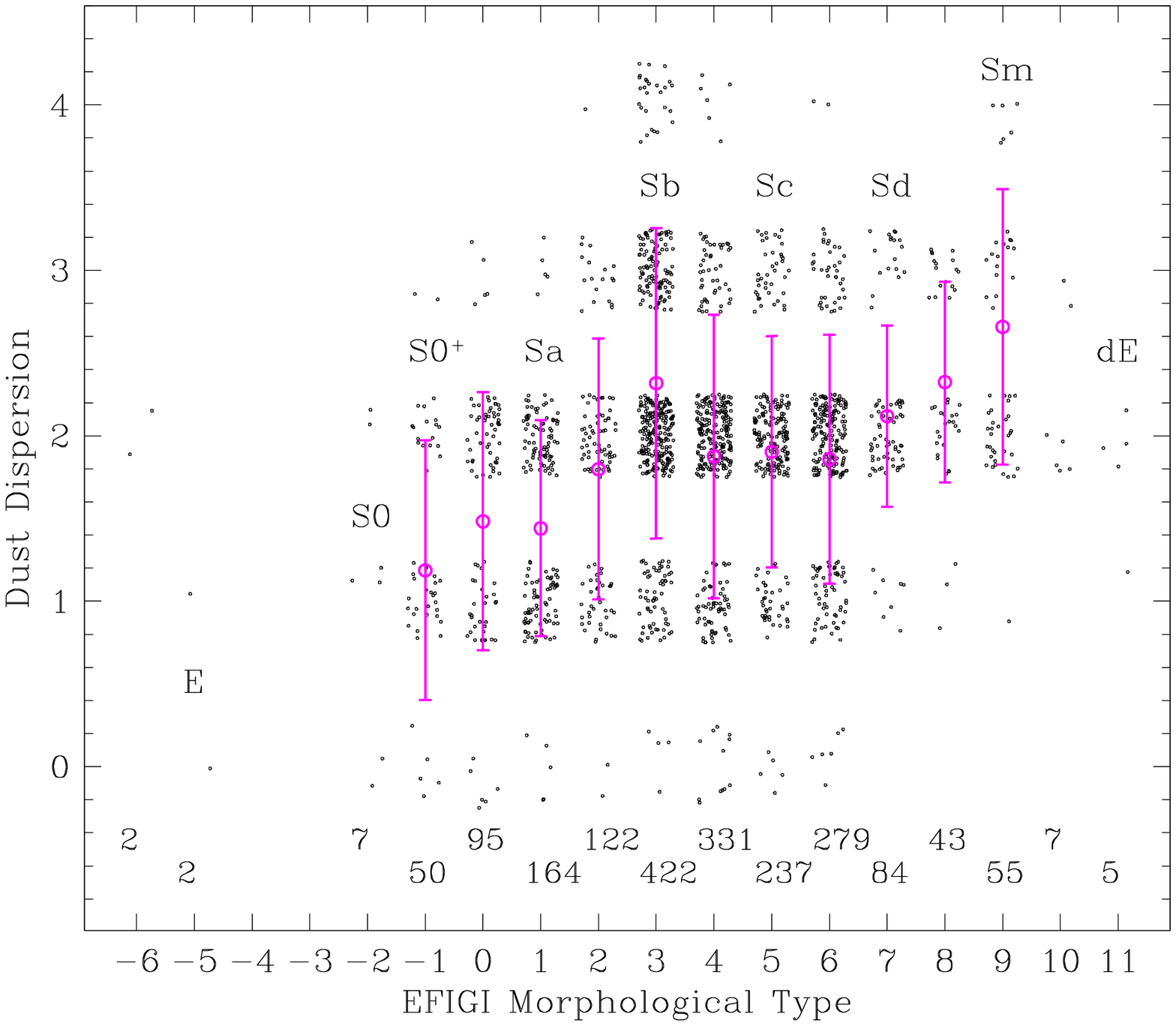,height=8cm,angle=0}}
\caption{Distribution of the \dust\ (left) and \disp\
  (right) attributes for the 4458 and 1905 EFIGI galaxies for
  which these attributes are defined (same presentation as in
  \fg\ref{emt_bt}). The left
  panel shows the large amounts of dust for early and intermediate
  spiral types (Sa to Scd), with a peak for Sb galaxies, and the
  lower values for late spiral (Sd to Sm) and Im types. The right
  panel shows the increasing \disp\ for spiral galaxies of
  later type.}
\label{emt_dust_disp}
\end{figure*}

\begin{table}
  \caption{Statistics of the EFIGI \dust\ attribute given as a percentage of galaxies with given attribute values}
\label{attrib_stat_dust}
\begin{center}
\begin{tabular}{lrrrrrr}                      
\hline  
\hline
EFIGI  &  \multicolumn{6}{c}{\tt Visible Dust}\\
 Type  &   0       &   1       &   2       &  3        &  4        & 1-2-3-4  \\
\hline
  cE   & 68$\pm$25 & 21$\pm$12 &  5$\pm$ 5 &  5$\pm$ 5 &         - & 32$\pm$15\\ 
  cD   & 96$\pm$20 &  4$\pm$ 3 &         - &         - &         - &  4$\pm$ 3\\ 
   E   & 97$\pm$ 9 &  2$\pm$ 1 &  1$\pm$ 1 &         - &         - &  3$\pm$ 1\\ 
 S0$^-$& 97$\pm$10 &  3$\pm$ 1 &         - &         - &         - &  3$\pm$ 1\\ 
  S0   & 85$\pm$ 9 & 11$\pm$ 2 &  4$\pm$ 1 &         - &         - & 15$\pm$ 3\\ 
 S0$^+$& 42$\pm$ 6 & 25$\pm$ 5 & 22$\pm$ 4 &  9$\pm$ 3 &  1$\pm$ 1 & 58$\pm$ 8\\ 
 S0a   & 24$\pm$ 4 & 26$\pm$ 4 & 32$\pm$ 5 & 16$\pm$ 3 &  2$\pm$ 1 & 76$\pm$ 8\\ 
  Sa   &  9$\pm$ 2 & 26$\pm$ 4 & 33$\pm$ 4 & 25$\pm$ 4 &  7$\pm$ 2 & 91$\pm$ 8\\ 
 Sab   &  4$\pm$ 1 & 41$\pm$ 5 & 30$\pm$ 4 & 18$\pm$ 3 &  8$\pm$ 2 & 96$\pm$ 9\\ 
  Sb   &  1$\pm$ 0 & 17$\pm$ 2 & 36$\pm$ 3 & 32$\pm$ 3 & 13$\pm$ 2 & 99$\pm$ 6\\ 
 Sbc   &         - & 29$\pm$ 3 & 47$\pm$ 4 & 19$\pm$ 2 &  4$\pm$ 1 &100$\pm$ 6\\ 
  Sc   &         - & 20$\pm$ 3 & 51$\pm$ 5 & 24$\pm$ 3 &  4$\pm$ 1 &100$\pm$ 8\\ 
 Scd   &  2$\pm$ 1 & 35$\pm$ 3 & 48$\pm$ 4 & 13$\pm$ 2 &  1$\pm$ 1 & 98$\pm$ 7\\ 
  Sd   & 21$\pm$ 3 & 50$\pm$ 5 & 25$\pm$ 3 &  4$\pm$ 1 &         - & 79$\pm$ 7\\ 
 Sdm   & 36$\pm$ 4 & 52$\pm$ 5 & 10$\pm$ 2 &  2$\pm$ 1 &         - & 64$\pm$ 5\\ 
  Sm   & 30$\pm$ 4 & 49$\pm$ 5 & 16$\pm$ 3 &  5$\pm$ 1 &         - & 70$\pm$ 7\\ 
  Im   & 74$\pm$ 7 & 23$\pm$ 3 &  3$\pm$ 1 &         - &         - & 26$\pm$ 4\\ 
  dE   & 86$\pm$15 &  7$\pm$ 3 &  6$\pm$ 3 &  1$\pm$ 1 &         - & 14$\pm$ 5\\ 
\hline
All spirals & 10$\pm$ 1 & 34$\pm$ 1 & 35$\pm$ 1 & 17$\pm$ 1 & 4$\pm$ 1 & 90$\pm$ 2 \\ 
All types   & 29$\pm$ 1 & 28$\pm$ 1 & 27$\pm$ 1 & 13$\pm$ 1 & 3$\pm$ 1 & 71$\pm$ 2 \\ 
\hline
\end{tabular}
\end{center}
{\it Note:} Null fractions are replaced by `` - '' for clarity.
\end{table}

We now examine another four attributes that further define the
properties of the spiral structure in spiral galaxies and play a role in
the visual definition of the morphological sequence: the presence of
dust (\dust), the patchiness of its distribution (\disp),
the significance of the scattered and giant HII regions (\floc\ and \spot\ respectively).

   \subsubsection{Dust                                                  \label{dust}}               

In \fg\ref{emt_dust_disp} we show the distribution of the \dust\ and \disp\
attributes as a function of EFIGI type; note that \disp\ is defined 
only for the 1905 galaxies with \dust\ $\ge 2$. In Table
\ref{attrib_stat_dust} we list the fractions of galaxy types 
for each value of the \dust\ attribute and for all values of \inc.
Contrary to the bar and ring attributes, dust is also visible
in highly inclined or edge-on galaxies and is even easier to detect
owing to a higher contrast; this is measured by
the systematically higher fractions of galaxies with \dust\ = 3-4 
for \inc\ = 3-4 compared to \inc\ = 1-2.

Dust appears to be frequent in galaxies: altogether, $90\pm2$\% of EFIGI 
spirals and $71\pm2$\% of all types show visually detectable dust.
The variations in \dust\ with type bear some ressemblance with those in \arm,
which peaks for Sc, but there are some significant differences. 
\fg\ref{emt_dust_disp}
shows a sharply increasing amount of dust for the intermediate lenticular 
types S0$^+$ and S0a ($58\pm8$\% and $76\pm8$\% host dust) up to the
early and intermediate type spirals (Sa to Scd), with a peak for Sb
types: more than 90\% of Sa to Scd types host dust, with 15 to 32\% at a
strong level; and Sb types have $13\pm2$\% of galaxies with \dust\ = 4. 
There is a much smaller relative amount of dust 
in late spirals (Sd to Sm) however, with a decreasing fraction of $79\pm7$\% to 
$70\pm7$\% galaxies altogether, less than 
$5\pm1$\% of these types at the strong level, and $25\pm3$\% to $16\pm3$\% at 
the moderate level. The overall frequency of \dust\ decreases to only 
$26\pm4$\% in Im galaxies. The low values of the \dust\ attribute for late 
spiral and irregular types is a combination of the lesser dust content of 
lower luminosity galaxies \citep{vandenbergh90b}, and the lower luminosity 
of late spirals and irregulars (de Lapparent \& Bertin 2011a, in prep{.}). 

The right panel of \fg\ref{emt_dust_disp} shows that the \disp\
is low for Sa galaxies, reaches a peak for Sb types followed by
a decrease and a``plateau'' for intermediate types (Sbc to Scd), then increases 
again for later spiral types. Strongly patchy dust is essentially observed in types
Sab to Sm.

\begin{figure*}
\centerline{\psfig{figure=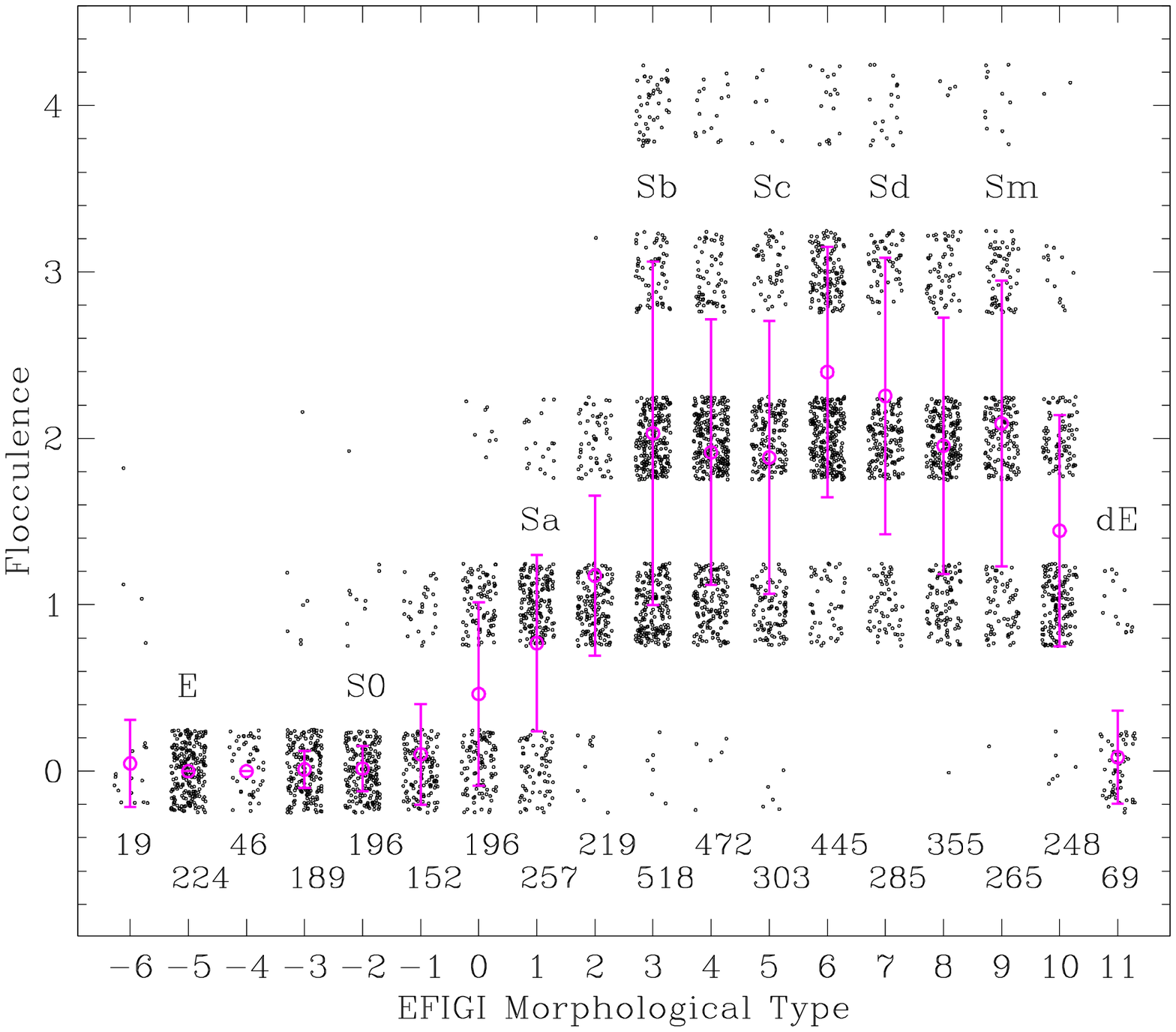,height=8cm,angle=0}
\psfig{figure=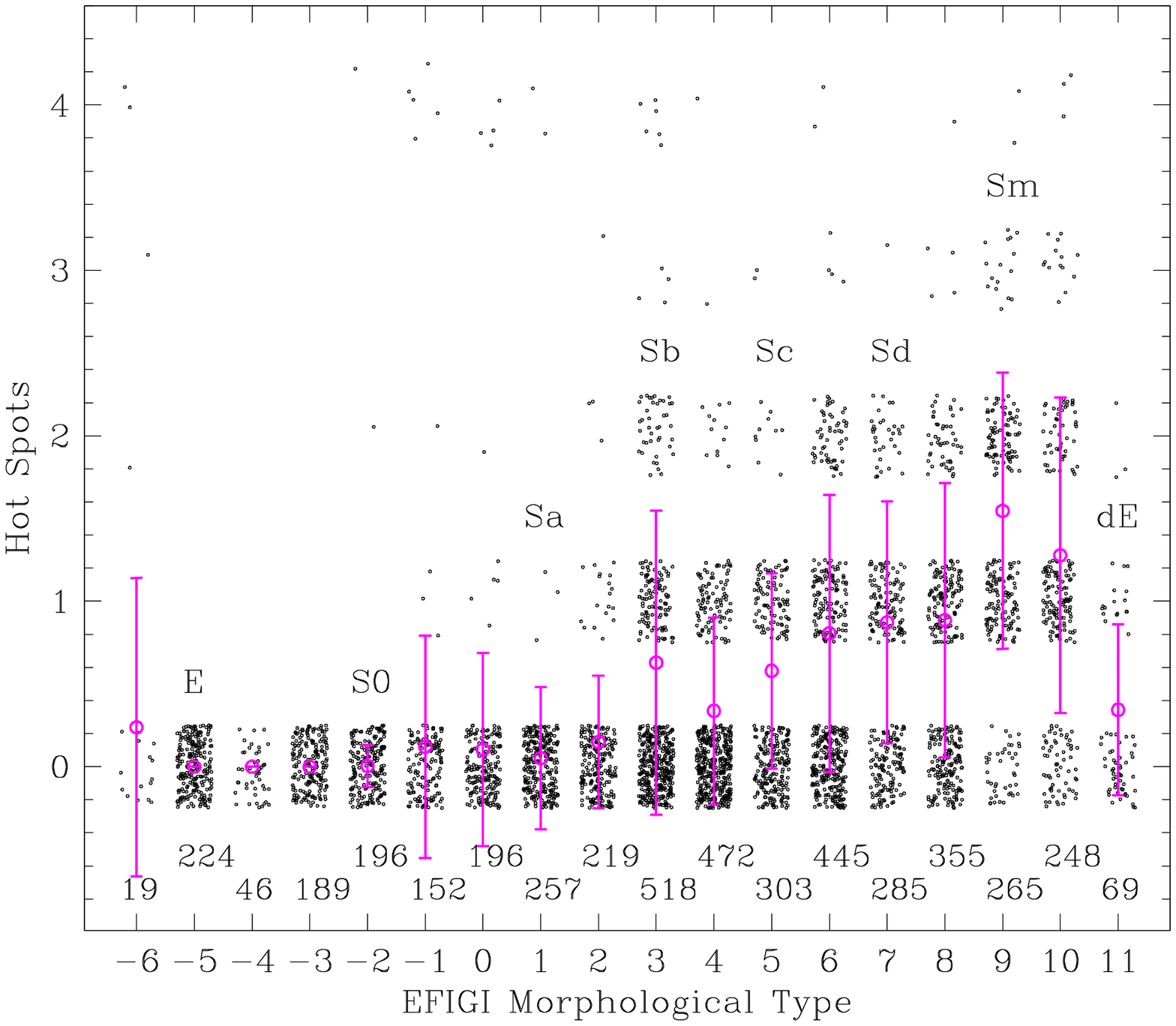,height=8cm,angle=0}}
\caption{Same as \fg\ref{emt_arm} for the \floc\ (left) and \spot\
  (right) attributes. The left panel shows the nearly constant
  level of \floc\ for all spiral types Sb and later and the
  lower values for Sa and Im galaxies. The right panel shows a
  nearly constant level of \spot\ for all spiral types from Sb to
  Sdm and an increased \spot\ attribute for Sm and Im types.}
\label{emt_floc_spot}
\end{figure*}

\begin{table}
  \caption{Statistics of the EFIGI \floc\ and \spot\ ``Texture'' attributes given as a percentage of galaxies with given attribute values}
\label{attrib_stat_text}
\begin{center}
\begin{tabular}{lrrrrrr}                      
\hline  
\hline
EFIGI   & \multicolumn{6}{c}{\tt Flocculence}\\
Type    &   0      &   1       &   2       &  3        &    4      & 1-2-3-4  \\
\hline
  cE   & 79$\pm$27 & 16$\pm$10 &  5$\pm$ 5 &         - &         - & 21$\pm$12\\ 
  cD   &100$\pm$21 &         - &         - &         - &         - &         -\\ 
   E   &100$\pm$ 9 &         - &         - &         - &         - &         -\\ 
 S0$^-$& 96$\pm$10 &  3$\pm$ 1 &  1$\pm$ 1 &         - &         - &  4$\pm$ 1\\ 
  S0   & 95$\pm$10 &  5$\pm$ 2 &  1$\pm$ 1 &         - &         - &  5$\pm$ 2\\ 
 S0$^+$& 82$\pm$10 & 18$\pm$ 4 &         - &         - &         - & 18$\pm$ 4\\ 
 S0a   & 50$\pm$ 6 & 46$\pm$ 6 &  4$\pm$ 1 &         - &         - & 50$\pm$ 6\\ 
  Sa   & 25$\pm$ 4 & 68$\pm$ 7 &  7$\pm$ 2 &         - &         - & 75$\pm$ 7\\ 
 Sab   &  4$\pm$ 1 & 72$\pm$ 8 & 23$\pm$ 4 &         - &         - & 96$\pm$ 9\\ 
  Sb   &  1$\pm$ 0 & 36$\pm$ 3 & 43$\pm$ 3 & 12$\pm$ 2 &  9$\pm$ 1 & 99$\pm$ 6\\ 
 Sbc   &  1$\pm$ 0 & 30$\pm$ 3 & 52$\pm$ 4 & 13$\pm$ 2 &  3$\pm$ 1 & 99$\pm$ 6\\ 
  Sc   &  2$\pm$ 1 & 32$\pm$ 4 & 49$\pm$ 5 & 15$\pm$ 2 &  2$\pm$ 1 & 98$\pm$ 8\\ 
 Scd   &         - & 10$\pm$ 2 & 57$\pm$ 4 & 29$\pm$ 3 &  4$\pm$ 1 &100$\pm$ 7\\ 
  Sd   &         - & 20$\pm$ 3 & 54$\pm$ 5 & 20$\pm$ 3 &  6$\pm$ 2 &100$\pm$ 8\\ 
 Sdm   &  0$\pm$ 1 & 26$\pm$ 3 & 57$\pm$ 5 & 15$\pm$ 2 &  1$\pm$ 1 &100$\pm$ 7\\ 
  Sm   &  0$\pm$ 1 & 23$\pm$ 3 & 48$\pm$ 5 & 24$\pm$ 3 &  4$\pm$ 1 &100$\pm$ 9\\ 
  Im   &  2$\pm$ 1 & 56$\pm$ 6 & 35$\pm$ 4 &  5$\pm$ 1 &  1$\pm$ 1 & 98$\pm$ 9\\ 
  dE   & 80$\pm$14 & 20$\pm$ 6 &         - &         - &         - & 20$\pm$ 6\\ 
\hline
All spirals &  3$\pm$ 0 & 32$\pm$ 1 & 46$\pm$ 1 & 15$\pm$ 1 & 4$\pm$ 1 & 97$\pm$ 2  \\ 
All types   & 23$\pm$ 1 & 29$\pm$ 1 & 34$\pm$ 1 & 11$\pm$ 1 & 3$\pm$ 1 & 77$\pm$ 2  \\ 
\hline
EFIGI   & \multicolumn{6}{c}{\tt Hot Spots}\\
Type    &   0     &   1     &   2     &  3       &  4      & 1-2-3-4  \\
\hline
  cE   & 79$\pm$27 &         - &  5$\pm$ 5 &  5$\pm$ 5 & 11$\pm$ 8 & 21$\pm$12\\ 
  cD   &100$\pm$21 &         - &         - &         - &         - &         -\\ 
  E    &100$\pm$ 9 &     - &         - &         - &         - &         -\\  
S0$^-$ &100$\pm$10 &     - &         - &         - &         - &         -\\ 
  S0   & 99$\pm$10 &         - &  1$\pm$ 1 &         - &  1$\pm$ 1 &  1$\pm$ 1\\ 
S0$^+$ & 94$\pm$11 &  2$\pm$ 1 &  1$\pm$ 1 &         - &  3$\pm$ 1 &  6$\pm$ 2\\ 
 S0a   & 95$\pm$10 &  3$\pm$ 1 &  1$\pm$ 1 &         - &  2$\pm$ 1 &  5$\pm$ 2\\ 
  Sa   & 98$\pm$ 9 &  1$\pm$ 1 &         - &         - &  1$\pm$ 1 &  2$\pm$ 1\\ 
 Sab   & 90$\pm$ 9 &  8$\pm$ 2 &  1$\pm$ 1 &  0$\pm$ 1 &         - & 10$\pm$ 2\\ 
  Sb   & 69$\pm$ 5 & 21$\pm$ 2 &  8$\pm$ 1 &  1$\pm$ 1 &  1$\pm$ 1 & 31$\pm$ 3\\ 
 Sbc   & 77$\pm$ 5 & 20$\pm$ 2 &  3$\pm$ 1 &  0$\pm$ 1 &  0$\pm$ 1 & 23$\pm$ 2\\ 
  Sc   & 64$\pm$ 6 & 32$\pm$ 4 &  3$\pm$ 1 &  1$\pm$ 1 &         - & 36$\pm$ 4\\ 
 Scd   & 57$\pm$ 4 & 29$\pm$ 3 & 13$\pm$ 2 &  1$\pm$ 1 &  0$\pm$ 1 & 43$\pm$ 4\\ 
  Sd   & 48$\pm$ 5 & 40$\pm$ 4 & 12$\pm$ 2 &  0$\pm$ 1 &         - & 52$\pm$ 5\\ 
 Sdm   & 47$\pm$ 4 & 38$\pm$ 4 & 14$\pm$ 2 &  1$\pm$ 1 &  0$\pm$ 1 & 53$\pm$ 5\\ 
  Sm   & 16$\pm$ 3 & 44$\pm$ 5 & 34$\pm$ 4 &  6$\pm$ 2 &  1$\pm$ 1 & 84$\pm$ 8\\ 
  Im   & 25$\pm$ 4 & 47$\pm$ 5 & 21$\pm$ 3 &  6$\pm$ 2 &  1$\pm$ 1 & 75$\pm$ 7\\ 
  dE   & 71$\pm$13 & 25$\pm$ 7 &  4$\pm$ 3 &         - &         - & 29$\pm$ 7\\ 
\hline
All spirals & 63$\pm$ 2 & 26$\pm$ 1 & 9$\pm$ 1 & 1$\pm$ 1 & 0$\pm$ 1 & 37$\pm$ 1\\ 
All types   & 69$\pm$ 2 & 21$\pm$ 1 & 8$\pm$ 1 & 1$\pm$ 1 & 1$\pm$ 1 & 31$\pm$ 1\\ 
\hline
\end{tabular}
\end{center}
{\it Note:} Null fractions are replaced by `` - '' for clarity.
\end{table}

   \subsubsection{Flocculence and hot spots                                    \label{spots}}               

Finally, the \floc\ and \spot\ attributes are useful for
quantifying the fraction of light in the scattered and giant HII
regions of galaxies. These are rare in elliptical, cD, and lenticular
galaxies and increase at types S0$^+$ and later. \fg\ref{emt_floc_spot} 
then shows a roughly constant
level of \floc\ and \spot\ for all spiral types from Sb to Sm, and Sb to Sdm. 
Then the \floc\ decreases for Im and dE galaxies, whereas 
the \spot\ attribute shows a marked increase for types Sm and Im, 
and again decreases for dE. 

Table \ref{attrib_stat_text} shows that
some level of flocculence is detected in more than 95\% of spiral
galaxies with types Sab to Sm, whereas hot spots occur in nearly one third
of Sb, Sbc, Sc, and Scd types, in half of Sd and Sdm types, 
in $84\pm8$\% of Sm types, and in $75\pm7$\% of Im types.
For Sa and Sab, hot spots occur in less than $2-10$\% of galaxies,
and in $\sim5-6$\% of S0$^+$ and S0a types. cE have a high fraction
of $21\pm12$\% of objects with evidence for hot spots, which are superimposed 
on the dense central parts of the objects. The high rate of dE with \spot\ $\ge1$ 
($29\pm7$\%) is due to the nucleated dS0 contained in this class. About half
of these nuclei seem associated with recent star formation as evidenced by
the marked blue colours in their vicinity (see also \citealp{paudel10}, who
recently showed that the stellar populations of these nuclei in Virgo galaxies
are young and metal-enhanced).

Altogether, $97\pm2$\% of spiral galaxies show evidence for flocculence, and
$37\pm1$\% for hot spots. The increase of the \floc\ and \spot\
attributes along the Hubble sequence is tightly related to
the gas content of the galaxies and the resulting star formation,
which both increase along the Hubble sequence. Because dust is produced
by star formation, the distribution of visible dust is somewhat related
to the flocculence, this comparison is complicated however by the fact that 
dust-enshrouded star formation is not detectable
in the optical. The similar trends in the distributions (left panels of
\fgs\ref{emt_dust_disp} and \ref{emt_floc_spot}) and statistics of
the \dust\ and \floc\ attributes may result from this relation.


\subsection{Specifying the Hubble sequence              \label{hubseq}}

We showed in the above sub-sections that all EFIGI morphological
attributes vary systematically with morphological type, thus
providing us with a quantitative description of the Hubble sequence 
in terms of each specific morphological
features. We summarise these relations here :

\begin{itemize}
\item The \bt\ ratio decreases regularly along the Hubble sequence, but
  there is a large $2\sigma$ spread of nearly five types for a given \bt\ value.
\item The regular increase in the \arm\ attribute results in an increase in the
  disk contribution along the Hubble sequence, with the latest spirals being 
  disk-dominated, whereas the early-type spirals are clearly bulge-dominated. 
  It is only in intermediate type spirals (Sb and Sbc) that most of the disk 
  light emission lies in the spiral arms.
\item The mean curvature of the spiral pattern regularly decreases (and the
  pitch angle increases) for later spiral types.
\item Bars are frequent among all galaxy types except E and dE, and
  are detected in 28 and 7\% of S0 and Im galaxies and in 22 to 56\%
  of spirals; the strongest bars lie in early spirals, with a peak
  frequency of strong and very strong bars in Sab.
\item Inner rings occur in 55\% or more of S0a, Sa, Sab galaxies, whereas
  outer rings decrease from 51\% to 15\% in these types;
  inner rings are also strongest for these three types and
  decrease in strength for earlier (S0) and later (spiral) types;
  outer rings are strongest in S0a types and occur essentially
  in early-type disks (S0, S0$^+$, S0a, Sa and Sab).
\item The outer pseudo-ring pattern occurs in approximately 10\% of barred
  S0a, Sa, Sab and Sb galaxies.
\item Non-disk galaxies and S0$^-$ show rare evidence for distortions
  in their symmetry. Nearly half of the types S0 to Sdm have perturbed
  symmetry at a low and moderate level, whereas the latest galaxy types, 
  Sm and Im, show evidence for frequent and significant perturbations 
  (79 and 71\%).
\item \dust\ is present in significant amounts in types S0 and later, 
  with a peak for Sb galaxies, and in decreasing
  amounts for spirals types approaching Sb; however, the \disp\ 
  regularly increases for types later than Scd.
\item There is an increase of the strength of the \floc\ and \spot\
  attributes from Sa to Sb, then both attributes remain constant between Sb and
  Sdm types.  Although the flocculence is stable for Sm and decreases
  for Im, the strength of the hot spots increases for both types.
\end{itemize}

These relations confirm the increasing pitch angle of the spiral arms as the 
primary criterion for defining the progression along spiral types in the Hubble 
sequence \citep{vandenbergh98}. Moreover, the increasing strength in the spiral arms
along the spiral sequence results from the decreasing disk contribution across the
full sequence from ellipticals through lenticulars and down to spirals. 

We also propose to interpret the original distinction between Sb to Sc types, 
based on the increasing continuity of the spiral arm design \citep{vandenbergh98}, 
as resulting from the simultaneous steep increase in the relative flux in the 
spiral arms and the decrease in the amount of dust from
types Sb to Sbc-Sc. In contrast, the increase in the fraction of light in the 
scattered and giant HII regions does not appear to directly participate 
in the establishment of the Hubble sequence. 

We emphasize that although the Hubble morphological types constitute a decreasing 
sequence of \bt\ ratio and an increasing sequence of disk contribution to the total 
galaxy flux, the large $2\sigma$ spread of approximately five types per value of \bt\ results 
from the prevalence of the spiral arm properties over the \bt\ ratio for defining 
the spiral sequence.


\section{Size and surface brightness along the Hubble sequence      \label{diam}}


\subsection{$D_{25}$ distribution                        \label{d25}}

Figure \ref{emt_d25kpc} shows the distribution of intrinsic isophotal
diameters calculated in \sct\ref{lim} as a function of EM-type.
The statistics for each value are overplotted: mean and dispersion, 
both using 3-sigma clipping to
exclude galaxies with possible erroneous redshifts and/or intrinsic $D_{25}$ estimate.
As expected, the large cD envelopes and the small cE and dE objects are at
the high and low ends of the size distribution, respectively. For all 
galaxy types, the \rms dispersion is in the range 
of 30-50\% of the mean value. The galaxies with high 
values of intrinsic $D_{25}$ compared the bulk of the objects for a
given EM-type (for exemple the Sm galaxy with $D_{25}\simeq52$ kpc
or the dE galaxies with $D_{25}\ge 15$ kpc) might have their
$D_{25}$ overestimated because of a faint extension of the isophotes at
large radii; indeed, the fitted disk scale lengths of
these objects are within the range of values for the other galaxies
of same EM-type (de Lapparent \& Bertin 2011a, in prep{.}).

\begin{figure}
\resizebox{\hsize}{!}{\includegraphics{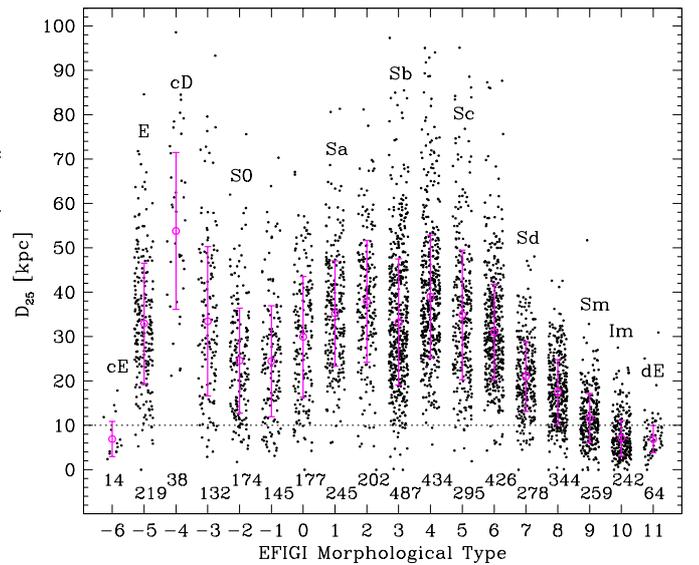}}
\caption{Distribution of intrinsic isophotal diameter in kpc for
  4196 EFIGI galaxies as a function of EFIGI morphological type. For
  each type, the weighted mean and \rms dispersion (with 3-sigma
  clipping) are plotted as a circle and vertical error-bar. The
  largest galaxies are the cD, Sbc and E, the smallest the cE, Im and
  dE. To guide the eye, a horizontal dotted line is drawn at 10 kpc.}
\label{emt_d25kpc}
\end{figure}

Moreover, the S0$^-$ have statistics comparable to E types,
whereas S0 and S0$^+$ galaxies have a lower mean intrinsic $D_{25}$, of
$\sim24$ kpc. This provides interesting support to the suggestion by
\citet{vandenbergh90} that some of the S0 and S0$^+$ (those with small
diameters) may not be intermediate between E and Sa. The Sa galaxies
also have a mean $D_{25}$ $\ga 30$ kpc, but with fewer galaxies at the
small diameter end compared to E types, which could be evidence that
the E galaxies are contaminated by small face-on lenticulars.  In
contrast, S0a types do appear to be intermediate in size between S0$^+$ and
Sa. The systematically smaller mean diameter for Sb galaxies compared to Sab
and Sbc must be checked by comparing the measured disk scale lengths
of the EFIGI spiral galaxies using the bulge+disk modelling (this is 
reported in de Lapparent \& Bertin 2011a, in prep{.}). 

Overall, \fg\ref{emt_d25kpc} shows a clear trend of steadily increasing mean
 diameter with EM-type for spiral galaxies from S0a up to Sab-Sbc galaxies.
Then, a gradual decrease occurs from Sc types to Im, with dE galaxies
having a similar size distribution as Im types. The ellipticals, lenticulars, 
and spirals earlier than Sd have mean diameters of 24 to 39 kpc, whereas 
galaxies from Sd to dE have mean diameters of 7 to 21 kpc. 

Moreover, small galaxies are few among Sa to Scd types:  less than 5\% of 
these EM-types have $D_{25}<10$ kpc (horizontal dotted line). S0$^-$ galaxy types
have slightly higher fractions of 7\% of galaxies with $D_{25}<10$ 
kpc, and S0 and S0$^+$ 
even higher fractions of 10\% and 15\% respectively. Despite their even smaller 
mean diameters of 21 kpc, Sd galaxies also have a low fraction of 7\% of such galaxies. 
In contrast, small spirals become frequent among Sdm and Sm types, 
with fractions of 17\% and 68\% of galaxies with $D_{25}<10$ kpc. 
Of course, cE, Im and dE galaxies have the bulk of their objects below
the $D_{25}=10$ kpc, whereas cD galaxies are all above this limit.

We have searched the EFIGI catalogue for dwarf spiral galaxies among types 
Sa to Scd. These are expected to be rare \citep{schombert95,sandage84}. 
We found two such objects using the criteria \arm
$\ge2$, and $g$ absolute magnitude fainter than -17: PGC0039483 and
PGC0040705, with magnitudes $-16.9$ and $-14.8$, and intrinsic
$D_{25}$ values of 5.7 kpc and 1.9 kpc. Both galaxies are located in the direction of
the Virgo cluster, with J2000 RA{.} and DEC{.} coordinates 
(184.499$^\circ$,6.654$^\circ$) and 
(186.635$^\circ$,12.610$^\circ$), but with significantly smaller recession
velocities of 791 km/s and 193 km/s (corrected for virgocentric infall).
Other names for PGC0039483 are NGC 4241 and IC 3115, whereas
PGC0040705 is also named NGC 4413 (NED also provides NGC 4407, but
it is not recognised in HyperLeda).

\fg\ref{dwarf_sp} shows
$irg$ colour images of both galaxies, obtained using the {\sc STIFF}
software\footnote{\tt http://www.astromatic.net/software/stiff}.
Both galaxies have a short bar (\bbar\ = 2) and weak flocculence 
(\floc\ = 1). PGC0039483 has open arms (\curv\ = 3), whereas those of 
PGC0040705 have an intermediate curvature (\curv\ = 2). 
The well designed spiral arms and small B/T ratio of PGC0039483 result in its
assigned Scd type in both the RC3 and the EFIGI catalogue.
In contrast, PGC0040705 was classified as Sab in the RC3, probably because of its 
tightly wound spiral arms, whereas in the EFIGI classification, it was assigned 
an Sc type that reflects its low B/T ratio and flocculence. 

By its small size, very faint absolute magnitude, and central nucleus, 
PGC0040705 recalls the spiral structure seen in some of the EFIGI
dE galaxies, which also have a nucleus and a bar.  \citet{barazza02} and
\citet{lisker06} showed that these objects are frequent among bright
dE galaxies in the Virgo cluster, in contrast with the
faint magnitude of PGC0040705. Because the EFIGI catalogue is
restricted to galaxies with several measurements of RC3 Hubble types,
and is only $\sim80$\% complete at bright magnitudes (Paper I),
some other nearby dwarf spirals might be missing.
A systematic search for dwarf spirals over the whole SDSS survey is
necessary to evaluate their true spatial frequency.

\begin{figure}
\resizebox{\hsize}{!}{\includegraphics{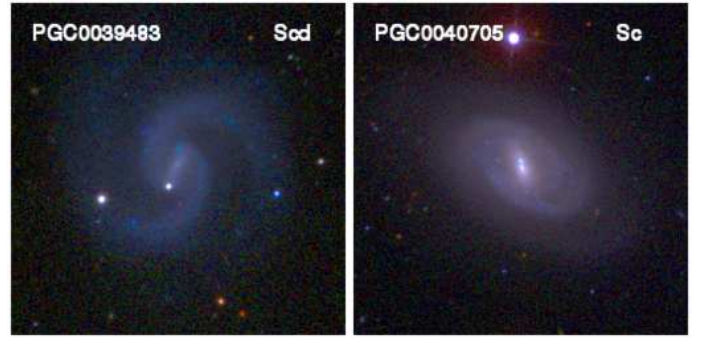}}
\caption{``True colour'' $irg$ images of the two dwarf spiral galaxies
  found in the EFIGI catalogue. These galaxies have redshifts
  0.0026385 and 0.0006414, absolute $g$ magnitudes of $-16.9$ and
  $-14.8$, and intrinsic $D_{25}$ values of 5.7 kpc and 1.9 kpc.}
\label{dwarf_sp}
\end{figure}


\subsection{Mean surface brightness                       \label{sb}}

\begin{figure}
\resizebox{\hsize}{!}{\includegraphics{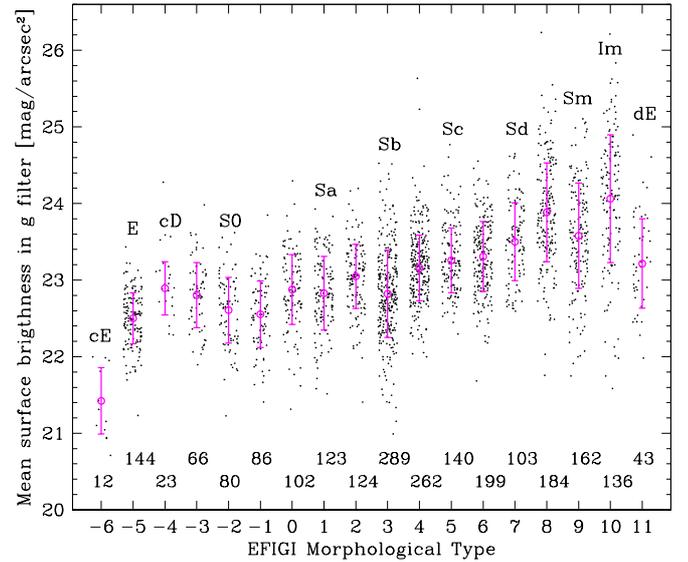}}
\caption{Distribution of mean surface brightness in the $g$-band
  within the $D_{25}$ diameter for the 2278 EFIGI galaxies 
  with \incc\ $\le2$ and \cont\ $\le1$
  as a function of EFIGI morphological type. For each
  type, the weighted mean and \rms dispersion (with 3-sigma clipping)
  are also plotted as a circle and vertical error-bar. The mean surface
  brightness is decreasing along the Hubble sequence from cE to Im
  types.}
\label{emt_sb25}
\end{figure}
\begin{figure}
\resizebox{\hsize}{!}{\includegraphics{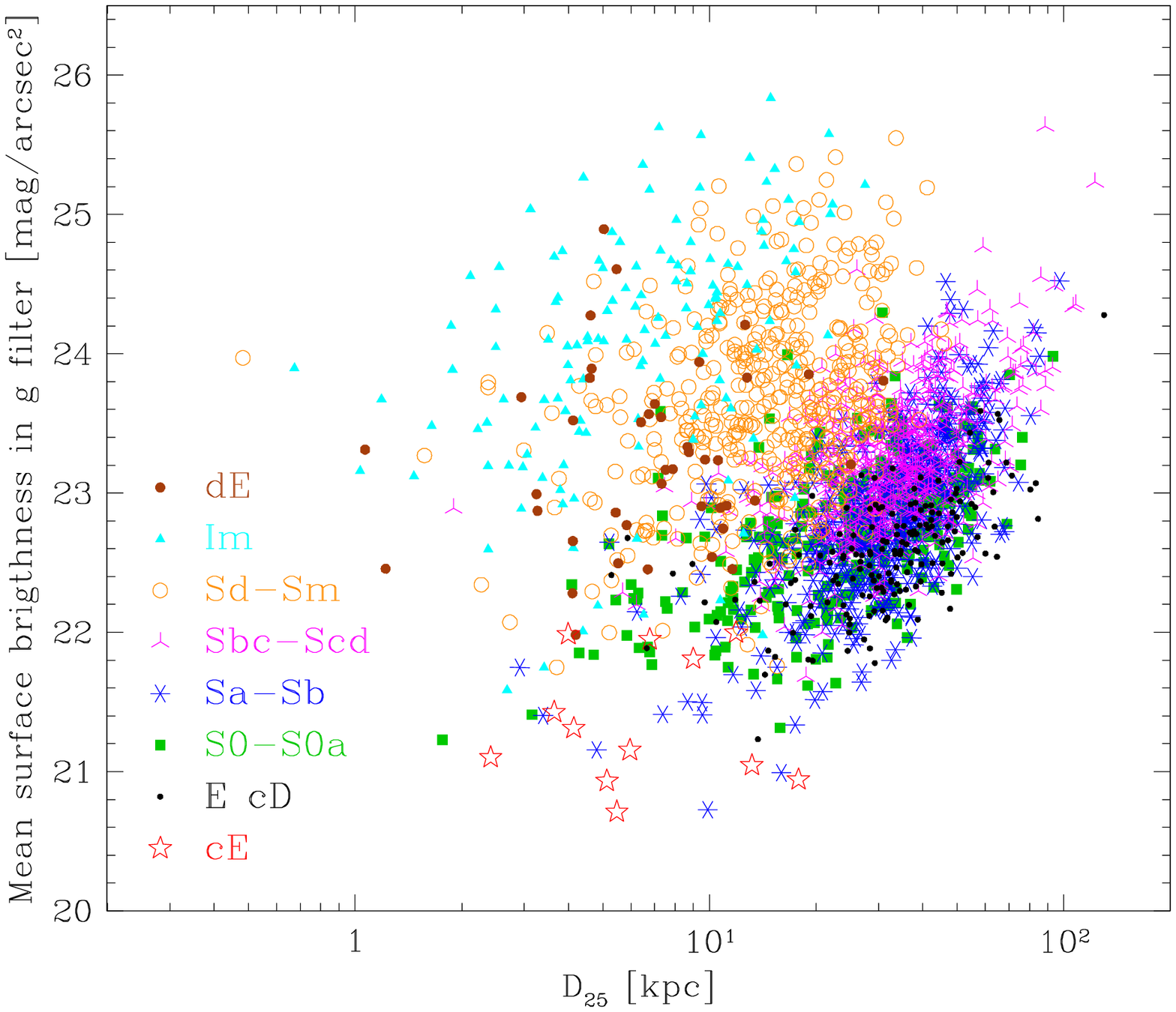}}
\caption{Distribution of apparent major isophotal diameter
  $D_{25}$ converted into kpc as a function of mean surface brightness
  in the $g$-band within the $D_{25}$ diameter for the 2278 EFIGI
  galaxies for which these parameters are defined and with \incc\
  $\le2$ and \cont\ $\le1$. Different symbols are used for different EM-types.
  This graph shows that galaxies with a surface brightness of dE but
  a more extended diameter could be detected in the EFIGI catalogue.}
\label{d25_sb25}
\end{figure}

One may wonder whether the small intrinsic diameters for the late
EM-type galaxies in \fg\ref{emt_d25kpc} is caused by some selection
effect at low surface brightness that characterises
late spiral, irregular and dE galaxies. We examined this effect by
estimating the mean surface brightness  (SB hereafter) for the EFIGI
galaxies, using again the apparent major isophotal diameter in the $D_{25}$
system. This estimate is much fainter than the central SB or that
calculated at the effective radius, but it still allows one to perform
an internal comparison among the various EM-types.  We only use the
2348 galaxies with small inclination (\inc\ $\le2$), and with no or negligible contamination
(\cont\ $\le1$): high contamination evidently
biases the SB towards bright values, whereas the SB of highly inclined
galaxies is biased towards faint values due to systematically higher
dust extinction of both the bulge and the disk than for face-on
galaxies \citep{driver07b}.

Figure \ref{emt_sb25} shows the distribution of mean SB using the
$g$-band magnitudes measured from the SDSS images by de Lapparent \& Bertin (2011a, in prep{.]) 
as a function of EM-type. This graph confirms the compactness of the cE galaxies,
which show the highest mean SB. If we exclude the cD galaxies, we observe
a dimming in  mean SB along the whole Hubble sequence, starting with E types.
Combined with the decrease in diameter for later spiral types shown in
\fg\ref{emt_d25kpc}, this results in the progressive decrease in absolute
luminosity of spiral galaxies along the Hubble sequence,
as shown by \citet{sandage85b} using the luminosity functions per morphological
type in the Virgo Cluster (see \citealp{lapparent03b}, for a review, and de Lapparent \& Bertin 2011b,
in prep{.} for luminosity functions in the EFIGI catalogue). 
The systematically brighter SB of Sb and Sm types compared to the monotonically
increasing curve relating SB to EM-type is examined by de Lapparent \& Bertin (2011a, in prep{.])
using SB estimates obtained from the bulge+disk modelling of each galaxy 
profile after convolution by the PSF.

A comparison of the mean SB of EFIGI galaxies with their intrinsic major
diameters, as shown in \fg\ref{d25_sb25}, allows one to evaluate how these
two galaxy characteristics could be biased by selection effects.
First, the SB limit at 25 mag/arcsec$^2$ mentioned in \sct\ref{lim}
seems to potentially affect mostly late spirals and irregulars, because
types earlier than Sd appear to be intrinsically limited in SB at 
$\sim24$ mag/arcsec$^2$ or brighter.

Moreover, the large galaxies (those with an intrinsic major diameter larger 
than 1 kpc) have an SB fainter than a limit determined by the sum of the
brightest absolute magnitude and $5\ln{D_{25}}$, where
$D_{25}$ is the intrinsic major diameter in kpc. This
yields the diagonal boundary in the lower right part of
\fg\ref{d25_sb25}.  There is also an
apparent boundary in the upper right part of the graph, indicating
that the EFIGI catalogue is deficient in large galaxies with very low SB
(Malin 1 type objects). It is not clear whether these objects are frequent
in the nearby universe and whether this boundary in \fg\ref{d25_sb25} is an
intrinsic or a selection effect; there is at present no complete census
of the galaxy distribution at very low SB (see \citealt{mcgaugh95} and 
\citealt{dalcanton97}).

At the small intrinsic diameter limit, one can see that the cE, dE, and
Im galaxies span the whole range of mean SB from bright to faint
values. The Im galaxies have a fainter SB tail than the dE galaxies
(see also \fg\ref{emt_d25kpc}). The scarcity of galaxies with a mean SB brighter
than 22 is due to the combination of the upper limit in galaxy
absolute luminosities (de Lapparent \& Bertin 2011b, in prep{.}) and the $D_{25}>1$ arcmin
limit on the angular size (see \fg\ref{d25_g}). 

Moreover, \fg\ref{d25_sb25} shows that galaxies larger than dE but
with the same mean SB could be detected in the EFIGI catalogue: these galaxies
would lie in the central part of the data cloud, which is mostly populated 
by the brightest of the Sd to Sm types, the smallest of the Sbc to Sbc types,
and a fair number of S0 to S0a and Sa to Sb galaxies. This
suggests that the small intrinsic diameter of dE galaxies in
\fg\ref{emt_d25kpc} is likely not due to a selection bias, but to an
intrinsic upper limit for this galaxy type. Similarly, \fg\ref{d25_sb25} 
indicates that larger galaxies of types Sd to Im could also be detected 
in the EFIGI sample. The small diameter of these galaxies therefore 
appears to be an intrinsic feature of these objects.


\section{Conclusions                  \label{conc}}

In addition to the surface brightness limit inherent to
all galaxy catalogues, the EFIGI sample is limited in apparent
diameter: following the RC3 selection, the majority of the EFIGI
galaxies have a major isophotal diameter larger than 1 arcminute in
the RC2 $D_{25}$ system \citep{rc2}.  This yields a dense sampling of
all Hubble types that differs from volume or magnitude limited
surveys by oversampling late spirals and irregulars, and therefore does not
reflect the real galaxy mix.  A strong advantage in this sampling
is that it provides us with large numbers of galaxies for each Hubble type
(except for the rare cE, cD and dE types), whose full diversity of
morphological characteristics can in turn be described using the EFIGI
attributes.  The various galaxy types are also widely distributed over
the redshift and absolute magnitude intervals of the survey. 

The high completeness rate of the EFIGI catalogue between the surface
brightness and apparent diameter limits does legitimate a statistical
analysis of the EFIGI attributes as a function of Hubble type. 
We thus examined the statistics of the
16 morphological attributes which were measured visually for each
of the 4458 EFIGI galaxies, in order to describe their various
components, their dynamical features, texture, contamination, and
environment. We estimated the frequency of each attribute and its
variations among the different Hubble types.

The analysis of EFIGI attributes confirms that the Hubble sequence is
an increasing sequence of bulge-to-total ratio and of disk
contribution to the total galaxy flux. Nevertheless, there is a large
$2\sigma$ spread of nearly five types for a given bulge-to-total ratio,
resulting from the prevalence of the spiral design in the progression
along the spiral sequence. The sequence is indeed characterized by a 
decreasing mean curvature of the spiral arms. We also propose that the decrease in 
visible dust from Sb to Sbc-Sc types combined with a 
steep increase in the strength of the spiral arms between these types
contributes to the emergence 
of the grand spiral design of Sc galaxies. The strength of the 
scattered and giant HII regions does not vary monotonically along the Hubble 
sequence, but along sub-portions of the sequence, with peaks and ``
plateaux'' for specific types (the \floc\ and \spot\ attributes
peak for Scd and Sm respectively); hence these attributes do not
seem to directly participate in the establishment of visual the Hubble sequence.
We also find that the deviation from a symmetric profile regularly increases 
along the Hubble sequence, but this is an incidental feature in the
definition of the sequence.

Dynamical features such as bars and inner rings are present in all types
of disk galaxies, and occur in $\sim40$\% and $\sim25$\% of them, 
with stronger bar/ring components in early-type spirals
and weaker components in late-type spirals.  Outer rings are nearly half 
as rare as inner rings, and both types of rings are tightly
correlated. Outer pseudo-rings occur in only $\sim10$\% of barred galaxies.

Owing to its $\sim80$\% photometric completeness with respect to the SDSS (Paper I),
with a preferential choice of galaxies with a reliable RC3 type, the spatial 
sampling of galaxy concentrations may suffer some bias, however. Systematic 
effects within each morphological type might exist, for example caused by
the combination of the variations from type to type in the \incc\ attribute 
and the fact that the RC3 Hubble types 
were easier to define for weakly inclined galaxies. The numerical 
values of the frequencies of the various EFIGI attributes as a function 
of type should therefore be used with caution. The detected 
systematic tendencies in the various attributes along the Hubble sequence 
are nevertheless likely to reflect intrinsic galaxy properties.

By cross-matching with the HyperLeda, NED, and SDSS catalogues
we obtained redshifts for 99.53\% of EFIGI galaxies ($z\la0.05$), which, 
combined with the $D_{25}$ measures, yield estimates of absolute 
major diameter and mean surface brightness in the SDSS $g$-band. 
We derive that the largest galaxies are cD, E, Sab, and Sbc
galaxies (20-50 kpc in $D_{25}$), and find a smooth decrease in size
along the Hubble sequence.  Late spirals are nearly twice smaller than
Sab-Sbc spirals, and Im, dE and cE types are indeed dwarf galaxies
(5-15 kpc in $D_{25}$). The lenticular galaxies are intermediate in
size (15-35 kpc in $D_{25}$). The surface brightness also smoothly
decreases along the Hubble sequence, but its selection limits within
the EFIGI sample do not exclude the detection of large late-type
galaxies.

In addition, we find that there are very few dwarf galaxies of types Sa to Scd,
in agreement with the observations of \citet{sandage84} in the
Virgo cluster: we find only two dwarf spirals in the EFIGI catalogue 
with well designed spiral arms.
We also notice an extension of Sb types to small $D_{25}$ diameter, which
must be examined in greater detail using a more reliable
estimate of size (de Lapparent \& Bertin 2011a, in prep{.}).
Profile fitting could also allow one to distinguish E galaxies from face-on S0, 
because these types might be confused in visual morphological
classifications \citep{vandenbergh90}.

If the present analysis of the EFIGI catalogue provides us with a description 
of the Hubble sequence in terms of each specific morphological feature, 
one step further is made by applying supervised learning tasks using 
``Support Vector Machines''. In this context, \citet{baillard08}
showed that one can determine the Hubble type from a
reduced number of EFIGI morphological attributes, which are, in
decreasing order of significance: the bulge-to-total luminosity
ratio, the strength of spiral arms, and the curvature of spiral arms. In that
analysis, the amounts of visible dust and flocculence play a less significant 
role in determining the Hubble type. These various results agree with 
the present analysis.

Two forthcoming articles provide examples of the
type of advanced morphological studies that can be performed using
the EFIGI database. A multi-component profile adjustment of each EFIGI galaxy
is carried out using the last version of {\sc SExtractor} \citep{bertin96,bertin10},
and the visual EFIGI attributes are used  
to calibrate and optimize the automatic profile fitting.
These analyses report on a bi-modality in the bulge
and disk colours derived from the fitted light profiles to EFIGI
galaxies (de Lapparent \& Bertin 2011a, in prep{.}), and present
the luminosity functions for the EFIGI sample as a function of morphological 
type and galaxy component (de Lapparent \& Bertin 2011b, in prep{.}).


\begin{acknowledgements}

  We are grateful to the referee, Ron Buta, 
  for his very useful comments.
  This work has been supported by grant 04-5500 (``ACI masse de
  donn\'ees'') from the French Ministry of Research.

  This research made use of the HyperLeda database
  (http://leda.univ-lyon1.fr), the VizieR catalogue access tool
  \citep{vizier} and the Sesame service at CDS (Strasbourg,
  France), and the NASA/IPAC Extragalactic Database (NED), which is
  operated by the Jet Propulsion Laboratory, California Institute of
  Technology, under contract with the National Aeronautics and Space
  Administration.

  This publication also made use of the Sloan Digital Sky Survey
  images and catalogues.  Funding for the SDSS and SDSS-II has been
  provided by the Alfred P. Sloan Foundation, the Participating
  Institutions, the National Science Foundation, the U.S. Department
  of Energy, the National Aeronautics and Space Administration, the
  Japanese Monbukagakusho, the Max Planck Society, and the Higher
  Education Funding Council for England. The SDSS Web Site is
  http://www.sdss.org/.  The SDSS is managed by the Astrophysical
  Research Consortium for the Participating Institutions. The
  Participating Institutions are the American Museum of Natural
  History, Astrophysical Institute Potsdam, University of Basel,
  University of Cambridge, Case Western Reserve University, University
  of Chicago, Drexel University, Fermilab, the Institute for Advanced
  Study, the Japan Participation Group, Johns Hopkins University, the
  Joint Institute for Nuclear Astrophysics, the Kavli Institute for
  Particle Astrophysics and Cosmology, the Korean Scientist Group, the
  Chinese Academy of Sciences (LAMOST), Los Alamos National
  Laboratory, the Max-Planck-Institute for Astronomy (MPIA), the
  Max-Planck-Institute for Astrophysics (MPA), New Mexico State
  University, Ohio State University, University of Pittsburgh,
  University of Portsmouth, Princeton University, the United States
  Naval Observatory, and the University of Washington.

\end{acknowledgements}


\bibliographystyle{aa} 
\bibliography{stat_hubseq}


\end{document}